\documentclass[12pt]{article}
\usepackage[ruled, vlined]{algorithm2e}
\usepackage{amsmath, amsfonts, amssymb, mathrsfs}
\usepackage{caption}
\usepackage{dcolumn}
\usepackage{filemod}
\usepackage{floatrow}
\usepackage{gensymb}
\usepackage{graphicx}
\usepackage{hyperref}
\usepackage{natbib}
\usepackage{setspace}
\usepackage{subcaption}
\usepackage{verbatim}
\usepackage{makecell}

\begin{document}

\title{Bayesian Statistical Inversion for High-Dimensional Computer Model
Output and Spatially Distributed Counts}
\author{Steven D. Barnett\thanks{Corresponding author
\href{mailto:sdbarnett@vt.edu}{\tt sdbarnett@vt.edu}, Department of Statistics,
Virginia Tech} \and Robert B. Gramacy\thanks{Department of Statistics, Virginia
Tech} \and Lauren J. Beesley\thanks{Statistics Group, Los Alamos National
Laboratory} \and Dave Osthus\footnotemark[3] \and Yifan Huang\thanks{ Center
for Space Plasma and Aeronomic Research, University of Alabama in Huntsville }
\and Fan Guo\thanks{Nuclear and Particle Physics, AstroPhysics and Cosmology,
Los Alamos National Laboratory} \and Daniel B. Reisenfeld\thanks{Space Science
and Applications Group, Los Alamos National Laboratory}}
\date{}

\maketitle

\begin{abstract}
Data collected by the Interstellar Boundary Explorer (IBEX) satellite,
recording heliospheric energetic neutral atoms (ENAs), exhibit a phenomenon that
has caused space scientists to revise hypotheses about the physical processes,
and computer simulations under those models, in play at the boundary of our
solar system.  Evaluating the fit of these computer models involves tuning
their parameters to observational data from IBEX. This would be a classic
(Bayesian) inverse problem if not for three challenges: (1) the computer
simulations are slow, limiting the size of campaigns of runs; so (2) surrogate
modeling is essential, but outputs are high-resolution images, thwarting
conventional methods; and (3) IBEX observations are counts, whereas most
inverse problem techniques assume Gaussian field data. To fill that gap we
propose a novel approach to Bayesian inverse problems coupling a Poisson
response with a sparse Gaussian process surrogate using the Vecchia
approximation.  We demonstrate the capabilities of our proposed framework,
which compare favorably to alternatives, through multiple simulated examples in
terms of recovering ``true'' computer model parameters and accurate
out-of-sample prediction. We then apply this new technology to IBEX satellite
data and associated computer models developed at Los Alamos National Laboratory.

\bigskip
\noindent \textbf{Key words:} Gaussian process, surrogate modeling,
Poisson, calibration, heliospheric science, IBEX, Vecchia approximation
\end{abstract}



\section{Introduction}
\label{sec:intro}

The National Aeronautics and Space Administration (NASA) launched the
Interstellar Boundary Explorer (IBEX) satellite in 2008 as part of their Small
Explorer program \citep{mccomas2009aIBEX} to deepen our understanding of the
heliosphere. The heliosphere is the bubble formed by the solar wind that
encompasses our solar system and acts as a barrier to interstellar space. Of
particular interest is the behavior at the edge of the heliosphere. Here,
highly energized hydrogen ions that make up the solar wind interact with
neutral atoms, occasionally undergo electron exchange, and become neutral
themselves. These energetic neutral atoms (ENAs) are unaffected by magnetic
fields and therefore travel in a ballistic trajectory.

\begin{figure}[ht!]
\centering
\includegraphics[scale=0.6,trim=0 0 85 0,clip=TRUE]{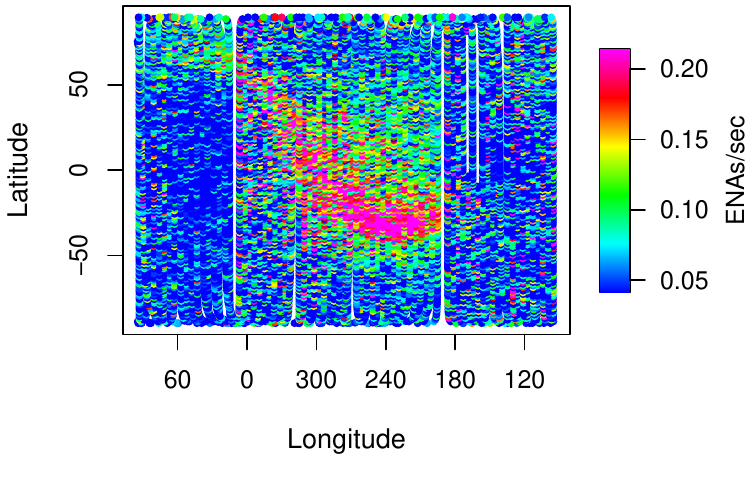}
\includegraphics[scale=0.6,trim=55 0 85 0,clip=TRUE]{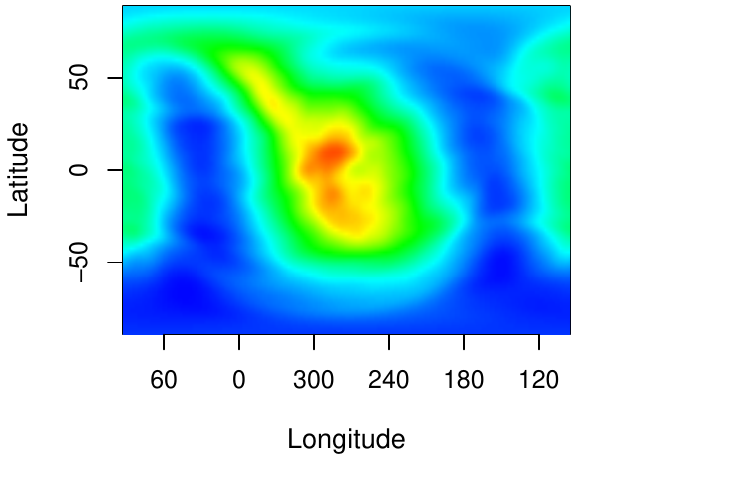}
\includegraphics[scale=0.6,trim=55 0 00 0,clip=TRUE]{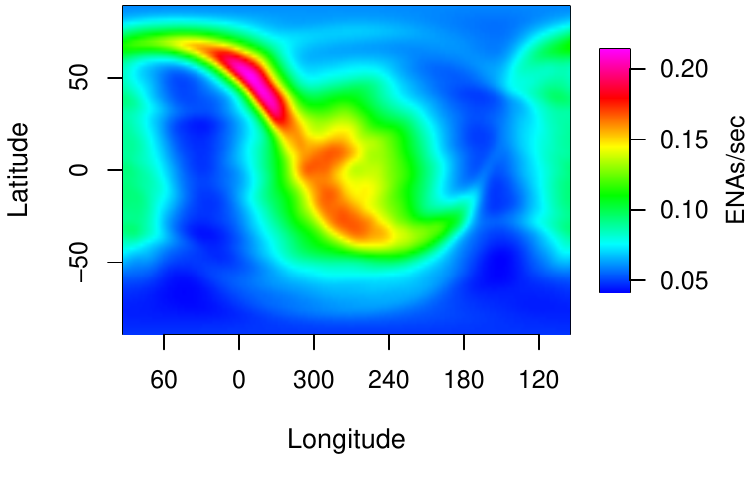}
\caption{Observed ENA rates detected by the
IBEX satellite (left) and output from two IBEX  simulations under
different parameter settings (middle, right).
\label{f:fig1}}
\end{figure}

Some ENAs eventually make their way to Earth and can be detected by an
instrument on the IBEX satellite called the IBEX-Hi ENA imager
\citep{funsten2009IBEXHiENA}. This apparatus records the energy level and
approximate location of origin for each ENA that enters the detector. IBEX's raw
collected data consist of ENA counts and exposure times for each area of the sky
at which the satellite points, providing sufficient information to estimate the
rate at which these particles are created throughout the heliosphere. An example
of the observed ENA rates detected by the IBEX satellite used to make
\textit{sky maps} (spatial maps of heliospheric ENA rates) can be found in the
left panel of Figure \ref{f:fig1}. Sky maps are integral to better understand
the heliosphere's properties and evolution.

One can think of the heliosphere as a ``boat'' moving through interstellar
space. At the outset of IBEX's mission, scientists expected to see ENAs being
generated throughout the heliosphere, with elevated rates at the front (nose)
and back (tail), much like the turbulent interaction with water at a boat's bow
and stern. And indeed, that belief was validated by IBEX data. At the nose and
tail of the heliosphere IBEX observed an increased number of collisions and
resulting electron exchanges between hydrogen ions and neutral particles. These
sources of ENAs, originating from the heliosheath (region between the
termination shock and the edge of the heliosphere), are identified as
\textit{globally distributed flux}, or GDF.

However, in a completely unanticipated finding, IBEX also recorded a thin
string of higher rates of ENAs \citep{fuselier2009IBEXribbon} wrapping around
the heliosphere. Scientists now refer to this phenomenom as the
\textit{ribbon}. This unique feature of a sky map is clearly visible in Fig.
\ref{f:fig1}. Since this discovery, space scientists have proposed several
theories attempting to describe the physical process that generates the ribbon
\citep{mccomas2014ibex, zirnstein2018role, zirnstein2019strong,
zirnstein2021dependence}.

Computer models encapsulating competing theories have been built, which generate
high-resolution (i.e, a high-dimensional response) synthetic sky maps (middle
and right panels of Figure \ref{f:fig1}). These simulations are extremely
sophisticated and involve many complex and expensive operations (e.g. solving
differential equations). Consequently, executing a single run of the computer
model at a specified set of model parameters can take hours, even when
leveraging parallel processing. This severely limits the number of unique runs
available to generate simulated sky maps in various settings.

Our work here focuses on two computer models provided to us by heliospheric
physicists involved in IBEX's mission: a GDF model proposed by
\citet{zirnstein2021heliosheath,zirnstein2025global,swaczyna2020density}; and a
ribbon-only model developed by \citet{huang2025numerical}. These simulators
take inputs that can be varied to modify either the GDF or the shape and
intensity of the ribbon. The ribbon-only model relies on two parameters of
interest, \textit{parallel mean free path} and \textit{ratio}, while we fix the
GDF model as constant and simply add it's response to the output of the ribbon
model. Dialing in good settings for these input parameters, in light of
observed ENA counts from IBEX, is instrumental in furthering theory development
and validation.

Space scientists at Los Alamos National Laboratory (LANL) want to know which
parameter settings best agree with IBEX observations. One way to do this is to
treat the problem as a Bayesian inverse problem
\citep{kaipio2011bayesian,stuart2010inverse,knapik2011bayesian}, obtaining a
posterior distribution over likely settings.  However, the exposure and count
data IBEX provides, along with computer simulated rates, present some unique
challenges: 1) computationally intensive models limiting simulation,
necessitating surrogate modeling; 2) high-dimensional model output (simulated
sky maps, each containing tens of thousands of pixels), thwarting conventional
surrogate modeling techniques; and 3) a non-Gaussian field response (Poisson
counts). It is worth noting that Bayesian posterior sampling, say via Markov
chain Monte Carlo (MCMC), compounds computational bottlenecks (2). We propose a
framework for Bayesian inverse problems that utilizes the Vecchia approximation
\citep{katzfuss2021general,scaledvecchiakatzfuss2022} with a Gaussian process
(GP) surrogate or emulator for high-dimensional computer simulations, and couple
that with a Poisson observational model to furnish posterior samples of unknown
parameters to these models given observed data.

Methods exist separately, in the literature, to address needs in each of the
aforementioned situations 1)--3), but we are not aware of any approaches in the
intersection. Table \ref{tab:prev_work} summarizes the features of the IBEX inverse
problem and reviews the necessary capabilities, alongside recent papers offering
partial solutions.  References provided are representative.  We acknowledge that
there are, in most cases, several works that may fit the bill.  Each row in the table
may be summarized as follows.

\begin{table}[ht!]
\centering
\renewcommand{\arraystretch}{1.5}
\begin{tabular}{|l|c|c|c|c|}
\hline
& Bayesian & \makecell{Non-\\Gaussian} & \makecell{Large-Scale \\ Simulation \\ Output} & \makecell{High-Dim \\ Response} \\
\hline
Our Contribution & \checkmark & \checkmark & \checkmark & \checkmark \\
\hline
\citet{kennedyohagan2001} & \checkmark & & & \\
\hline
\citet{higdoncalib2008} & \checkmark & & & \checkmark \\
\hline
\citet{gramacy2015calibrating} & & & \checkmark & \\
\hline
\citet{grosskopfcountcalib2020} & \checkmark & \checkmark & & \\
\hline
\end{tabular}
\caption{Check marks indicate that functionality exists within the cited paper and/or
accompanying software. To save space, only one citation is listed for each
row.}
\label{tab:prev_work}
\end{table}

\citet{kennedyohagan2001} offer the canonical computer model calibration
framework, which provides a fully Bayesian approach to estimate the posterior
distribution of model parameters and improve prediction at new, unobserved field
locations. However, their methodology is limited to Gaussian field observations
and small, simulator output. \citet{higdoncalib2008} and SEPIA
\citep{gattiker2020lanl}, its more recent incarnation, also employ Bayesian
methods while accounting for computer model output that is functional. But SEPIA
is somewhat rigid, only considers Gaussian field observations, does not scale
well with added model runs, and requires the user to specify a number of basis
functions to represent the simulator response. \citet{gramacy2015calibrating}
consider computer model calibration with large-scale output from a computer
experiment, necessitating an approximation. But their implementation is not
Bayesian and the scale of simulator data is still orders of magnitude less than
IBEX. \citet{grosskopfcountcalib2020} provide an extension of the Kennedy \&
O'Hagan (hereafter referred to as KOH) framework to non-Gaussian data that
maintains a Bayesian approach. But they are limited to small amounts of computer
model data and do not consider higher dimensional responses. In contrast, our
approach checks all the boxes to achieve the goal of IBEX space scientists and
therefore applies to a wider class of problems than previous work.

Our integrated framework for Bayesian inverse problems overcomes these obstacles
together in an all-in-one solution, as illustrated in the top row of Table
\ref{tab:prev_work}. We adopt the approach of generalized computer model
calibration, adapting to IBEX's count data with a Poisson likelihood and
employing MCMC to garner posterior samples of model parameters, which determine
the underlying mean of our Poisson model, given observed data. These samples
allow us to estimate the full joint posterior distribution of the unknown
parameters. We access predicted ENA rates through a GP surrogate or emulator fit
to a limited set of expensive, high-dimensional IBEX computer simulations to
more thoroughly explore the domain of possible sky maps that serves as the basis
for the observed ENA counts. As previously stated, MCMC exacerbates the
computational burden of making surrogate predictions, necessitating not only a
sparse GP surrogate, but one that can swiftly generate repeated sky map
proposals at each MCMC iteration. We find the Vecchia approximation to be
extremely effective in this regard. Lastly, we propose a more intuitive
technique to modeling high-dimensional vectors, treating each pixel representing
an ENA rate as a scalar response of interest. This differs from previous
approaches to functional computer model output, but enables us to leverage
recent advances in sparse GP surrogate modeling. Additionally, using this
strategy efficiently scales both to increased dimensionality of the response and
to larger sizes of training sets.

Our paper is laid out as follows. First, Section \ref{sec:review} reviews
methods that form the foundation for our work, namely Bayesian inverse problems
and Gaussian process surrogate models, and introduces our framework for solving
the inverse problem in the context of non-Gaussian field observations. We also
give a brief description of the methodology and implementation behind SEPIA,
the current ``status quo'' in Bayesian inverse problems for high-dimensional
computer model output. Section \ref{sec:vecchia} describes our use of the
Vecchia approximation to stretch the capacity of a GP surrogate beyond small
simulator training runs. In Section \ref{sec:ibex_data} we return to our
motivating application, illustrating the effectiveness of our framework in
discovering the true, underlying model parameters on synthetic data and apply
our methods to the raw IBEX satellite data. To conclude, we discuss any
extensions and future work in Section \ref{sec:discuss}. Our implementation may
be found at \url{https://github.com/lanl/ibex-bayesian-inverse} along with code
reproducing all included examples.



\section{Bayesian inverse problems}
\label{sec:review}

We study IBEX through the lens of observational data $Y^F$ ($F$ for field
measurement) and a computer implementation of a physical model $m(u, X)$,
where parameters $u$ are unknown and $X$ represents the spatial locations
where the simulation output is evaluated.  These may be the locations $X^F$
that correspond to field measurements $Y^F$, or a grid of predictive locations
supporting high resolution visuals.  Given prior information about $u$, we
want to know which $u$'s lead to realizations of $m(\cdot, \cdot)$ that could
explain $Y^F$, i.e., the posterior for $u \mid Y^F$.

This is a classic Bayesian inverse problem
\citep{kaipio2011bayesian,stuart2010inverse,knapik2011bayesian}. Eq.~\ref{eq:inv_bayes_mod} 
provides the typical hierarchical model used in this
setting, generally with Gaussian field data $Y^F$.
\begin{align}
Y^F &\sim \mathcal{N}_{n_F}\!\left(\mu\!\left(X^F\right),
\Sigma\!\left(X^F\right)\right) \nonumber  \\
\mu(X) &= m(u, X) \label{eq:inv_bayes_mod} \\
\Sigma(X) &\sim \pi\left(\Sigma\right) & \mbox{e.g.,} \quad \Sigma^{-1} & \sim \mathrm{Wishart}((\rho V)^{-1}, \rho)
\nonumber \\
u &\sim \pi(u) & \mbox{e.g.,} \quad u &\sim \mathrm{Unif}[0,1]^p \nonumber
\end{align}
One big difference with Bayesian inverse problems, as opposed to ordinary
Bayesian inference, is that the parameters $\mu(X^F)$ and $\Sigma(X^F)$
defining the distribution of $Y^F$ are not of direct interest. Rather, we are
interested in settings $u$ that define
$\mu(X)$ and $\Sigma(X)$ through $m(u, X)$.  For example, two different
settings of $u$ were used in the IBEX simulator to generate the center and
right panels of Figure \ref{f:fig1}. Colors indicate output $\mu(X)$ values on
a dense grid $X$ of longitudes and latitudes. We are interested in learning
which $u$ are likely to have generated the recorded $Y^F$ values in the left
panel, observed at locations where the IBEX satellite was pointed ($X^F$) to
collect ENAs.

\begin{figure}[ht!]
\centering
\includegraphics[scale=0.55,trim=20 0 30 0,clip=TRUE]{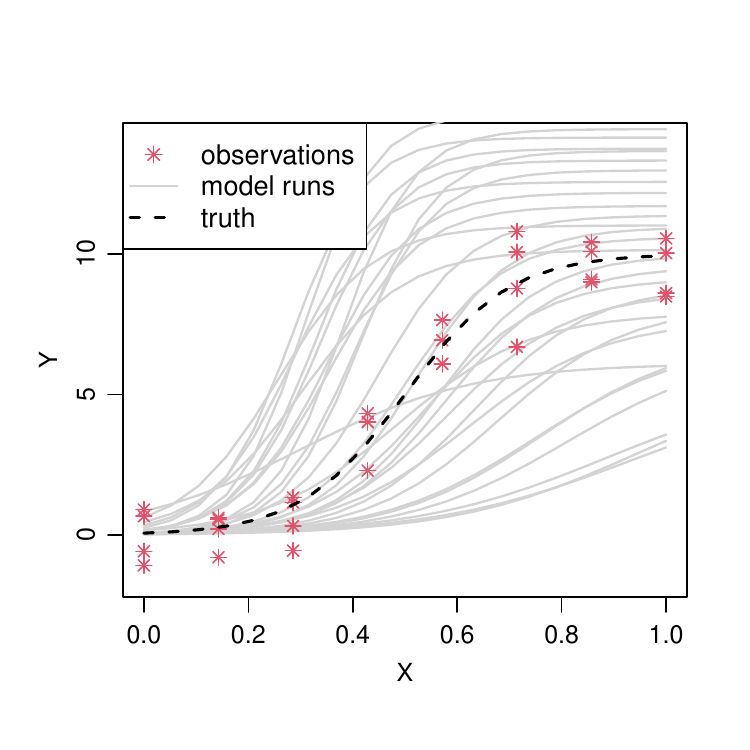}
\includegraphics[scale=0.55,trim=59 0 30 0,clip=TRUE]{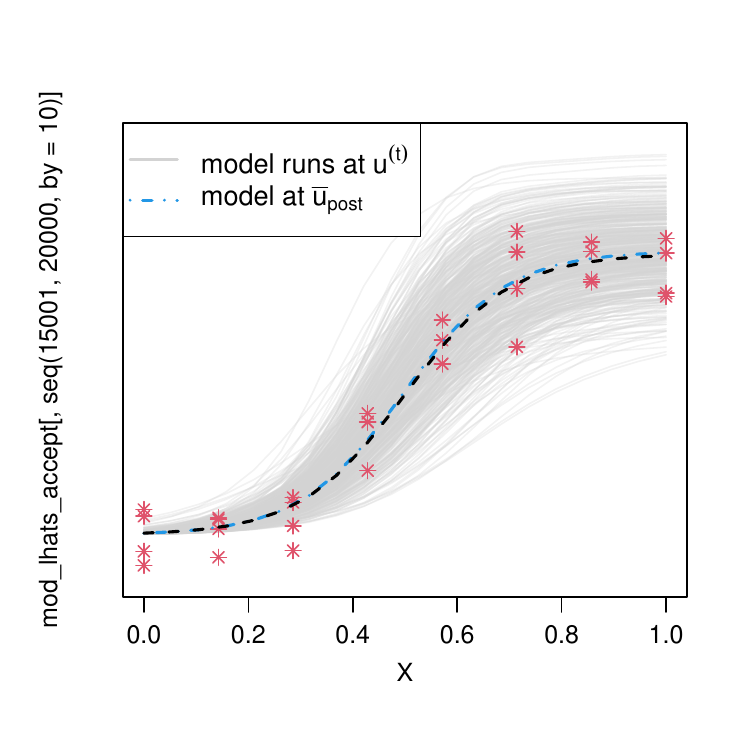}
\includegraphics[scale=0.55,trim=16 0 30 0,clip=TRUE]{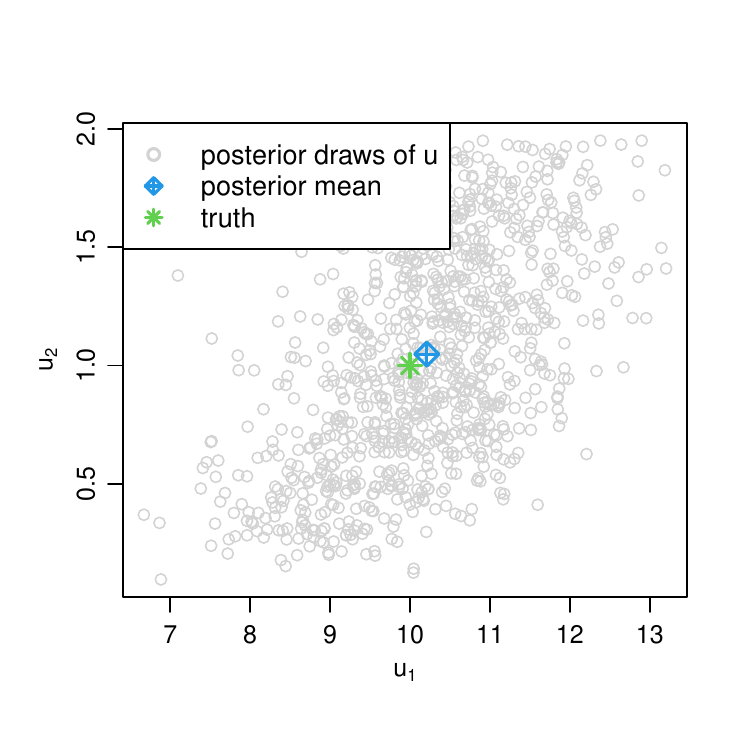}
\caption{A simple inverse problem. Field observations and a small sample of
computer model runs (left), model evaluations at posterior draws and mean
(middle), and posterior distribution of model parameters (right).
\label{f:toy_calib}}
\end{figure}

To help fix ideas, consider the simpler example introduced in the left panel
of Figure \ref{f:toy_calib}. Red stars indicate noisy field measurements,
$Y^F$, observed in replicate as $y_i \stackrel{\mathrm{iid}} \sim
\mathcal{N}(m(u^\star, x_i), \sigma^2)$ for a ``true'' unknown $u^\star =
(u_1^\star, u_2^\star)^\top$ and uniform $x_i \in X^F$.  In this simple
example, the computer model $m(\cdot, \cdot)$ was used to generate the data.
In practice, the model represents a caricature using known or best-guess
physics. Each light gray line is a realization of $m(\cdot, \cdot)$ for a
different $u$ on a fine grid of spatial locations $X$. The bold-black-dashed
line is $m(u^\star, X)$ provided as a reference.

The middle panel of the figure shows runs of our virtual model evaluated at posterior
draws of $u$ obtained under Metropolis sampling using a multivariate
normal likelihood whose mean $\mu(X^F)=m(u^{(t)}, X^F)$ and whose covariance $\Sigma(X^F) \sim \sigma^2
\mathbb{I}$, where $\sigma^2$ is marginalized out under a reference prior
$\pi(\sigma^2) \propto 1/\sigma^2$. Although we display these runs on a
high-resolution grid $X$, the sampler only uses evaluations of $m(u^{(t)}, X^F)$ in
likelihood evaluations. The posterior samples of $u$ thus obtained are shown
in the right panel of the figure.  In this toy example, it is evident that the
Metropolis sampler successfully concentrates posterior mass around the true
value of $u$.

The literature on Bayesian inverse problems involves methods that work
predominantly in this way, particularly as regards a Gaussian likelihood. We
aim to provide a method that achieves the same goal, but for unknown parameters
$u$ that underlie observed IBEX satellite data (left panel of Figure
\ref{f:fig1}), whose observations are counts under variable exposure. Our first
contribution, which is albeit rather straightforward, is to depart from that
literature by using a Poisson likelihood.

\subsection{Generalized Bayesian inverse problems}
\label{sec:gen_bayes_inv}

Our IBEX observations $Y^F$ are counts under exposure $e(X^F)$, which is at odds
with the canonical Gaussian likelihood in the Bayesian inverse problem setup
introduced earlier. We wish to model a count $y \sim \mathrm{Pois}(\lambda, e)$
where $f(y|\lambda, e) = \frac{(\lambda e)^y \mathrm{exp}\{-\lambda e\}}{y!}$
and $E(y) = \lambda e$. We thereby extend Eq.~(\ref{eq:inv_bayes_mod}) as
follows:
\begin{align}
Y^F &\sim \mathrm{Pois}(\lambda(X^F), e(X^F)) & \lambda(X) &= m(u, X) \nonumber
\label{eq:pois_inv_bayes_mod} \\ u &\sim \pi(u) & \mbox{e.g.,} \quad u &\sim \mathrm{Unif}[0,1]^p,
\end{align}
where $e(X^F)$ represents the duration (in seconds) the satellite points at
specific locations in the sky. $\lambda(X^F)$ denotes the rate at which ENA
particles are generated for rows (latitude/longitude pairs) in $X^F$. Sampling
from the posterior could proceed via Metroplis as in Algorithm \ref{alg:gibbs}.
For the IBEX field data, we evaluate a Poisson likelihood given an exposure
time for each observation and a mean surface determined by a simulator
$m(\cdot,\cdot)$ at proposed values of $u$, although we believe it would be
straightforward to accommodate other observational models.
\begin{algorithm}[ht]
\DontPrintSemicolon
Initialize $u^{(0)}$. \\
Set $\lambda(X^F)^{(0)} = m(u^{(0)}, X^F)$. \\
Compute and store $\mathcal{L}(Y^F|\lambda(X^F)^{(0)}, e(X^F)) =
\prod_{i=1}^{n_F} f(y_i|\lambda_i(X^F)^{(0)}, e_i)$, \\
\For{$t = 1, \dots, T$}{
  Propose new model parameters: $u^{(t)} \sim q(u^{(t)} \mid u^{(t-1)})$, \\
  \vspace{0.2em}
  Evaluate simulator at field locations and $u^{(t)}$: $\lambda(X^F)^{(t)} = m(u^{(t)}, X^F)$,
  \\
  \vspace{0.2em}
  Calculate and save $\mathcal{L}(Y^F|\lambda(X^F)^{(t)}, e(X^F)) =
  \prod_{i=1}^{n_F} f(y_i|\lambda_i(X^F)^{(t)}, e_i)$, \\
  \vspace{0.2em}
  Compute the acceptance ratio: $\alpha = \frac{\mathcal{L}\left(Y^F \mid
  \lambda(X^F)^{(t)}, e(X^F)\right) \pi\left(u^{(t)}\right) q\left(u^{(t-1)}
  \mid u^{(t)}\right)}{\mathcal{L}\left(Y^F \mid \lambda(X^F)^{(t-1)},
  e(X^F)\right) \pi\left(u^{(t-1)}\right) q\left(u^{(t)} \mid
  u^{(t-1)}\right)}$, \\ Accept or reject $u^{(t)}$. }
\caption{MCMC (Metropolis-within-Gibbs) for Poisson Bayesian inverse problems}
\label{alg:gibbs}
\end{algorithm}

\begin{figure}[ht!]
\centering
\includegraphics[scale=0.55,trim=20 0 30 0,clip=TRUE]{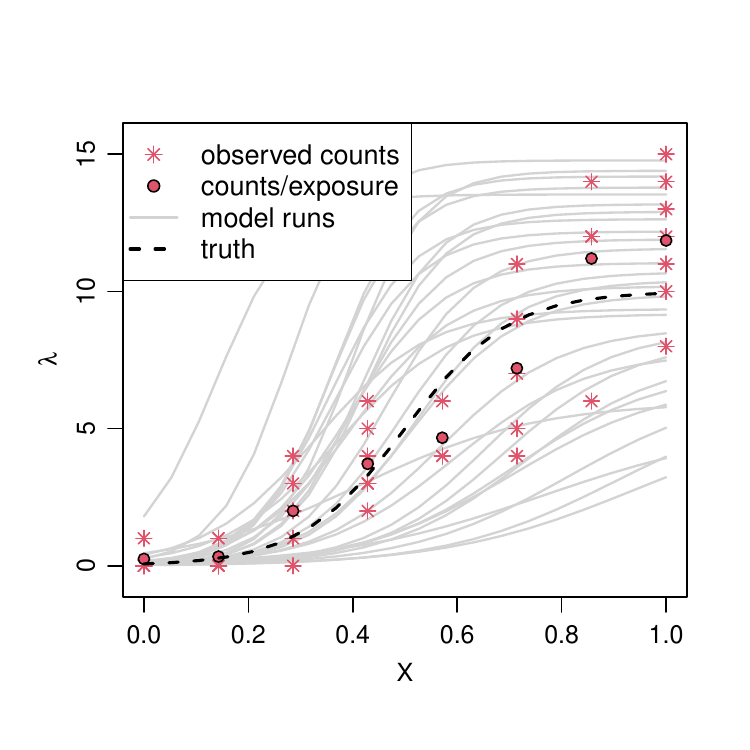}
\includegraphics[scale=0.55,trim=59 0 30 0,clip=TRUE]{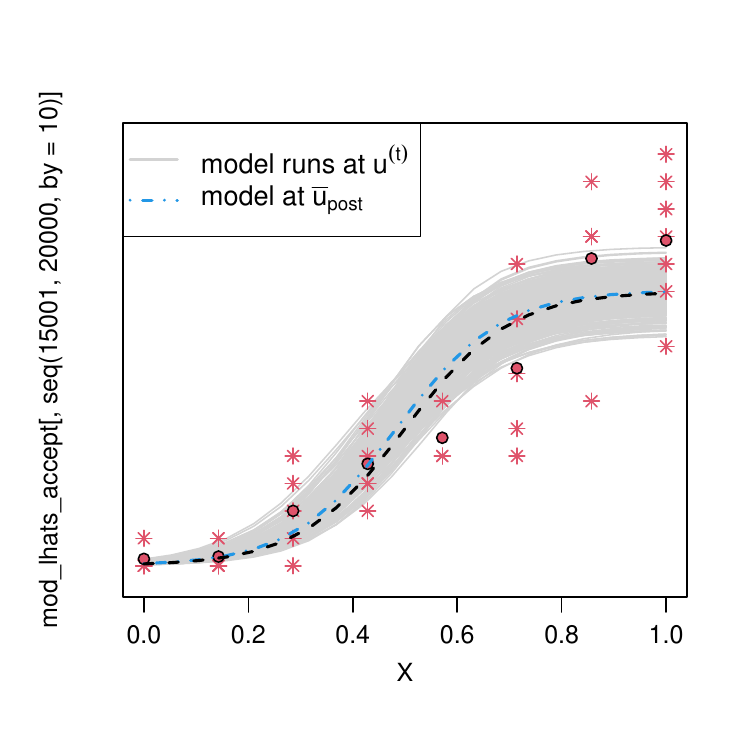}
\includegraphics[scale=0.55,trim=16 0 30 0,clip=TRUE]{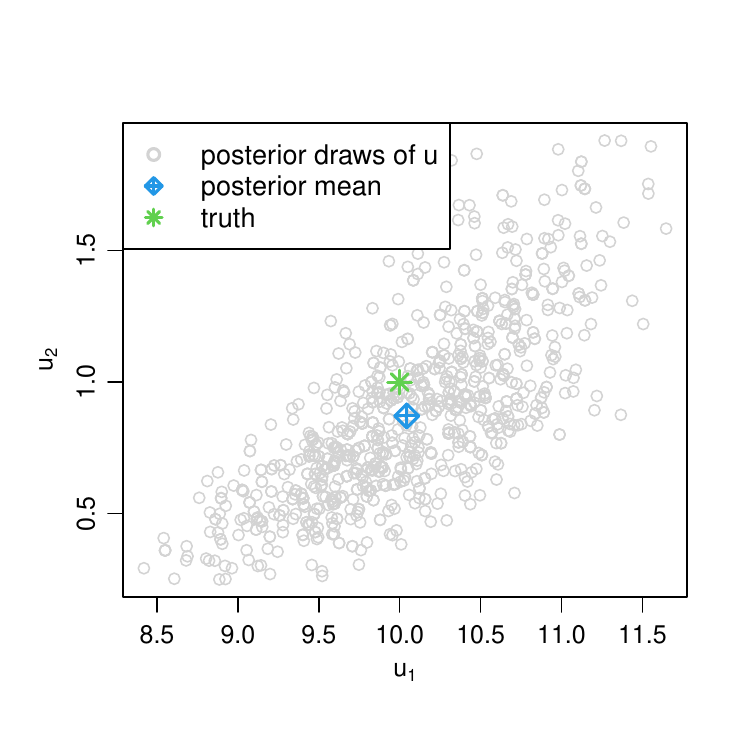}
\caption{An inverse problem for a Poisson process. Counts observed in the field
and a sample of computer model runs (left), model evaluations at posterior
draws and mean (middle), and posterior distribution of model parameters
(right).
\label{f:toy_calib_pois}}
\end{figure}

Using Algorithm \ref{alg:gibbs}, we can build on the toy problem we introduced
in Figure \ref{f:toy_calib} and align it more closely with the IBEX data.
Instead of observing a mean process $\mu(X^F)$ with random, Gaussian noise,
suppose we are given counts, drawn from a Poisson process with $\lambda(X^F) =
m(u^\star, X^F)$, as shown in the left panel of Figure \ref{f:toy_calib_pois}.
Computer model runs $m(\cdot,\cdot) = \lambda(X)$ produce different mean
surfaces given unique values of $u = (u_1, u_2)$. Observed counts are shown as
red stars. Red circles indicate the sum of all counts at one input location
(i.e, a single row of $X^F$) divided by the sum of their corresponding exposure
times $e(\cdot)$. These points are the actual data we receive. Results from
running Algorithm \ref{alg:gibbs} are shown in the middle and right panels of
Figure \ref{f:toy_calib_pois}. Similar to the results in Figure
\ref{f:toy_calib}, it is clear that the sampler searches a broad swath of the
model input space but hones in on the region where the true parameter value
lies.

In each of these toy examples we have made the assumption that we have unlimited
access to simulator runs (i.e, for each proposed combination of model
parameters, we can quickly get a response from the computer simulation). But
that is rarely the case, presenting another challenge to the typical Bayesian
inverse problem apparatus. For instance, each run of the IBEX model $m(\cdot,
\cdot)$ is a real challenge, requiring hours, even when evaluated on a
supercomputer.  That simply cannot be embedded in a Metropolis sampler
requiring thousands of iterations to converge.  While it's rather conventional
to deploy a surrogate model \citep[e.g.][]{gramacy2020surrogates} in this
setting, trained on a design of space-filling $u$-values, doing so effectively
is challenged by the number of rows (i.e. unique measurement locations) in
$X^F$.  We considered multiple surrogate models developed to accomodate such
scenarios (e.g. GPs via inducing points \citep{banerjee2008gaussian}, local
approximate GPs \citep{gramacy2014lagp}), but found them unable to manage the
scale of the IBEX data. Herein lies our main methodological contribution, using
a modern sparse approach to surrogate modeling in a (Poisson) Bayesian inverse
problem setting. The remainder of this section outlines the standard surrogate
modeling toolkit, as a jumping-off point toward explaining both our main
comparative (Section \ref{sec:gp_approx}) and our main methodological
contribution.

\subsection{Gaussian process surrogates for computer experiments}
\label{sec:gp_surr}

Evaluating the likelihood $Y_F$ in Algorithm \ref{alg:gibbs} may proceed via
surrogate when {\em in situ} evaluation directly on $m(\cdot, \cdot)$ is too
slow. Suppose we have a vector of observations $Y^M$ ($M$ for computer model) at
input locations $X^M$ from a campaign of runs of $m(\cdot, \cdot)$ that may or
may not coincide with settings in $X^F$. We prefer a small design of
space-filling $u$-values ($U$), paired via Cartesian product with $X^F$ from the
field experiment or a predictive grid $X$, e.g., $X^M = X \otimes U$, depending
on the situation. Two such runs, i.e., for two rows $u^\top \in U$ and a dense
grid $X$, are shown in Figure \ref{f:fig1}.

Given a corpus of runs $(Y^M, X^M)$, a surrogate synthesizes this information
in order to extend it to new inputs.  Many surrogates for computer
simulations are based on Gaussian processes
\citep[e.g.,][]{rasmussen2003gaussian}, where modeling boils down to
specifying a multivariate normal (MVN) for all responses.  This way,
prediction is simply a matter of MVN conditioning rules, or {\em kriging}
\citep{matheron1963principles}, to derive the distribution of unknown
(testing) output given known (training) values. Specifically, let $Y^M
\sim \mathcal{N}_{n_M}(\mu(X^M), \Sigma(X^M))$, where $\mathcal{N}_{n_M}$
indicates an $n_M$-variate MVN. Then, predicting at a new location $x$ (think
proposed $u$-value and paired row in $X^F$) follows $Y(x) \mid Y(X^M) \sim
\mathcal{N}(\hat{\mu}(x), \hat{\sigma}^2(x))$, where
\begin{align}
\hat{\mu}(x) &= \mu(x) + \Sigma(x, X^M) \Sigma(X^M)^{-1} (Y^M - \mu(X^M)) \label{eq:mvn_eq}\\
\hat{\sigma}^2(x) &= \Sigma(x, x) - \Sigma(x, X^M) \Sigma(X^M)^{-1} \Sigma(X^M, x). \nonumber
\end{align}
Although a generic mean function is specified in Eq.~(\ref{eq:mvn_eq}), it is
common to set $\mu(\cdot)=0$ assuming pre-centered responses. In this zero-mean
context, all the action takes place in the covariance function $\Sigma(x, x')$,
which is usually specified as inversely proportional to distances between inputs
$x$ and $x'$. We use a form known as the separable Mat\'ern kernel
\citep{stein1999interpolation} in our work, a conventional choice, parameterized
by lengthscales $\theta$ and scale $\tau^2$. For details, see \citep[][Chapter
5]{gramacy2020surrogates}. GPs provide a flexible, nonparametric regression
tool. In addition to enjoying canonical status as surrogates for computer
simulation experiments, they are utilized in geospatial statistics
\citep{cressie2011statistics}, time series \citep{roberts2013gaussian}, and
optimization \citep{jones1998efficient}.

GP surrogates are extremely effective. To illustrate, recall Figures
\ref{f:toy_calib} and \ref{f:toy_calib_pois} in Section \ref{sec:gen_bayes_inv},
where limited samples of computer model evaluations ($n < 50$) are shown. For
these toy examples, we ``assumed'' we had unlimited access to the simulator. In
reality, we fit a GP surrogate to the model runs in the left panels. Therefore,
``model runs'' at posterior draws of $u$ in the center panels are not in fact
evaluations of the model, but predictions from a fitted GP surrogate. Even so,
the Metropolis sampler used those accurate (and fast) predictions to
successfully recover the true model parameters $u^\star$.

One downside to working with MVNs whose dimension grows with the dimension of
the data they are trained on, $n_M$, is that building $n_M \times n_M$
matrices, and inverting (or otherwise decomposing) them, e.g., for
Eq.~(\ref{eq:mvn_eq}) requires storage and computation that is quadratic and
cubic in $n_M$, respectively.  This limits $n_M$ to a few thousand at most in a
Bayesian MCMC setting, necessitating approximations for GPs that curb this
prohibitive cost. With the 1d example in Figure \ref{f:toy_calib_pois}, we have
$n_M=20 \times 30$, combining $X$ and $U$, which is manageable. But for IBEX,
the simulated sky maps like those in the middle and right panels of Figure
\ref{f:fig1} are in 2d (under the hood the output is in spherical coordinates
$x$, $y$, and $z$) and results in ${>}16{,}000$ locations (or pixels) for each
$u$-value. Combine that with $n=66$ unique $u$-values used in runs available to
us and our response vector will have greater than one million elements. Even if
we used a smaller LHS space-filling design of size $n=15$ over the model
parameters, the response vector would still consist of over 200,000 points. In
either case, an ordinary GP could not be fit and provide predictions in a
reasonable amount of time, prompting the next topic.

\subsection{Approximations}
\label{sec:gp_approx}

Most of our modeling effort in this paper involves circumventing the $O(n_M^3)$
computational bottleneck associated with GPs. The prevalence of methods
addressing this issue has grown significantly in the past decade as increased
access to computing resources leads to problems with larger and larger training
datasets. These methods include, but are not limited to, fixed-rank kriging
\citep{cressie2008fixed}, inducing points \citep{banerjee2008gaussian},
compactly supported kernels \citep{kaufman2011efficient}, divide-and-conquer
\citep{gramacy2015local}, and nearest neighbors \citep{datta2016hierarchical,
wu2022variational}. Most of these are general enough to be used in a variety of
different fields and applications. However, there is a method tailor-made for
our Bayesian inverse problem setting when the likelihood (for $Y_F$) is
Gaussian.

Specifically in the context of (Gaussian) inverse problems,
\citet{higdoncalib2008} faced computational bottlenecks much like the IBEX
simulator: expensive simulations limiting runs and high-dimensional output in
the form of images, shape description, time series, etc. Even with today's
computing resources, it would be prohibitive to fit a GP surrogate model on such
large data. \citeauthor{higdoncalib2008}'s key insight involved treating $X^F$
(or gridded $X$) -- i.e., the fixed reference set that a simulator realization
is evaluated on for each candidate parameter $u$ -- as a feature of {\em
output}, rather than conventionally as an input. As a result, model and field
data share the same set of measurement locations, that is, $X = X^F$. In this
way, $X^M$ remains small, including only the unique combinations of unknown
parameters $u$.

Then, rather than keep track of all of those outputs $Y^M$, indexed over $X^F$
or $X$, dimensionality may be drammatically reduced via principal components
analysis (PCA). Specifically, they proposed representing the vector output of a
simulator model $m(\cdot,\cdot)$ as the sum of $n_k$ basis functions $k_1, k_2,
\ldots , k_{n_k}$, each scaled by a corresponding weight $w_i(u)$: $m(u, X)
\approx \sum_{i=1}^{n_k}  k_i \cdot w_i (u)$, where $w_i(u) \sim \mathcal{N}( 0,
\Sigma_{w_i})$ and $n_k \ll n_X$. Weights $w_i(u)$ are each modeled as
realizations of a zero-mean GP. By not directly modeling the high-dimensional
response $m(\cdot,\cdot)$ and it's associated dense grid $X$, requisite
decompositions $\Sigma_{w_i}$ avoid the cubic bottleneck in computation when
scaling to large $n$ and can be completed in a reasonable amount of time.

While clever, this is one of those easy-to-say hard-to-do situations. The
implementation is complex both mathematically, and in code.  We are aware of
only one library implementation, originally written in {\sf MATLAB} and now
deployed in {\sf Python} \citep{gattiker2020lanl}.  We found it difficult to
appropriate that setup for our own use in a Poisson likelihood context.
Additionally, restricting the simulated and field data to work on the same
reference grid was not possible with IBEX's orbital data locations (Figure
\ref{f:fig1}).

Fortunately, GP approximation has come a long way in the last
twenty or so years. \citeauthor{higdoncalib2008} specifically lamented not having
access to such technology, necessitating a more complex approach. It is no
longer necessary to play the trick of moving a reference grid to the output
space.  A GP (approximation) may model $Y^M$ via all inputs $X^F$ and $U$ via
$X^M$, as we shall demonstrate next.  The result is at once more modern, faster,
and better compartmentalized for use in novel contexts, such as our own.



\section{Vecchia Approximated GPs for IBEX Simulations}
\label{sec:vecchia}

In 1988, \citeauthor{vecchia1988estimation} proposed an approximation for
evaluating a multivariate normal density. The Vecchia approximation, say in the
context of GP modeling with the likelihood for $Y^M$ (i.e. $\mathcal{L}(Y^M)$),
relies on the ability to write any joint probability as a cascade of univariate
conditionals, which for GPs are Gaussian (\ref{eq:mvn_eq}).
\begin{align}
\mathcal{L}(Y^M)&=\prod_{i=1}^{n_M} \mathcal{L}(Y^M_i \mid Y^M_{g(i)})
\hspace{0.3em} \mbox{ where } g(1)=\emptyset \hspace{0.3em} \mbox{ and }
g(i)=\{1, 2, \ldots, i-1\} \nonumber \\
&\approx \prod_{i=1}^{n_M}\mathcal{L}(Y^M_i \mid Y^M_{h(i)}) \hspace{0.3em} \mbox{
where } h(i) \subset \{1, 2, \ldots, i-1\} \hspace{0.3em} \mbox{ and } |h(i)| \ll n_M
\label{eq:vecchia}
\end{align}

\citeauthor{vecchia1988estimation}'s critical observation was that one could
approximate this function by reducing the size of the conditioning set $g(i)$ to
a much smaller size $m$, as specified by the user. This induces sparsity in the
precision matrix (e.g. $\Sigma(X^M)^{-1})$ from Eq.~\ref{eq:mvn_eq}, potentially
dramatically speeding up the evaluation of the likelihood for large values of
$n_M$. \citet{katzfuss2021general} recently made this approximation popular by
working out its representation in matrix form and furnishing automatic $h(i)$
creation via nearest neighbors and publicly available code, allowing for broad
use in a variety of different fields where GPs are employed, particularly
spatial statistics and computer experiments
\citep{guinness2018permutation,katzfuss2020vecchia,katzfuss2021general,zhang2022multi,sauer2023vecchia}.

The Vecchia approximation takes an operation of order $O(n_M^3)$ down to $O(n_M
m^3)$, motivating small values of $m$. Logically, as the size of $h(i)$
decreases, the evaluation of the likelihood speeds up, but the quality of the
approximation is reduced. On the flip side, as one increases $m$, execution
time grows, but the degree of accuracy improves. This continues until $n=m-1$,
when the likelihood is no longer an approximation. For the IBEX simulator data,
we find a value of $m=25$ to be more than sufficient.

Several implementations of the Vecchia approximation exist, but we employ one
that has risen to the top in terms of predictive performance and computational
thriftiness in the context of computer experiments, that is, the Scaled Vecchia
approximation introduced by \citet{scaledvecchiakatzfuss2022}. We refer you to
their manuscript for a more in-depth explanation of their work, but the primary
draw for using it as a surrogate for the IBEX simulator is the open source
implementation and packaging that leverages modern sparse matrix libraries and
parallel processing. Code that we directly rely on is readily accessible at
\url{https://github.com/katzfuss-group/scaledVecchia}.

\subsection{Surrogate Sky Maps}
\label{sec:surr_skymaps}

\begin{figure}[ht!]
\centering
\includegraphics[scale=0.5, trim=5 0 30 0,clip=TRUE]{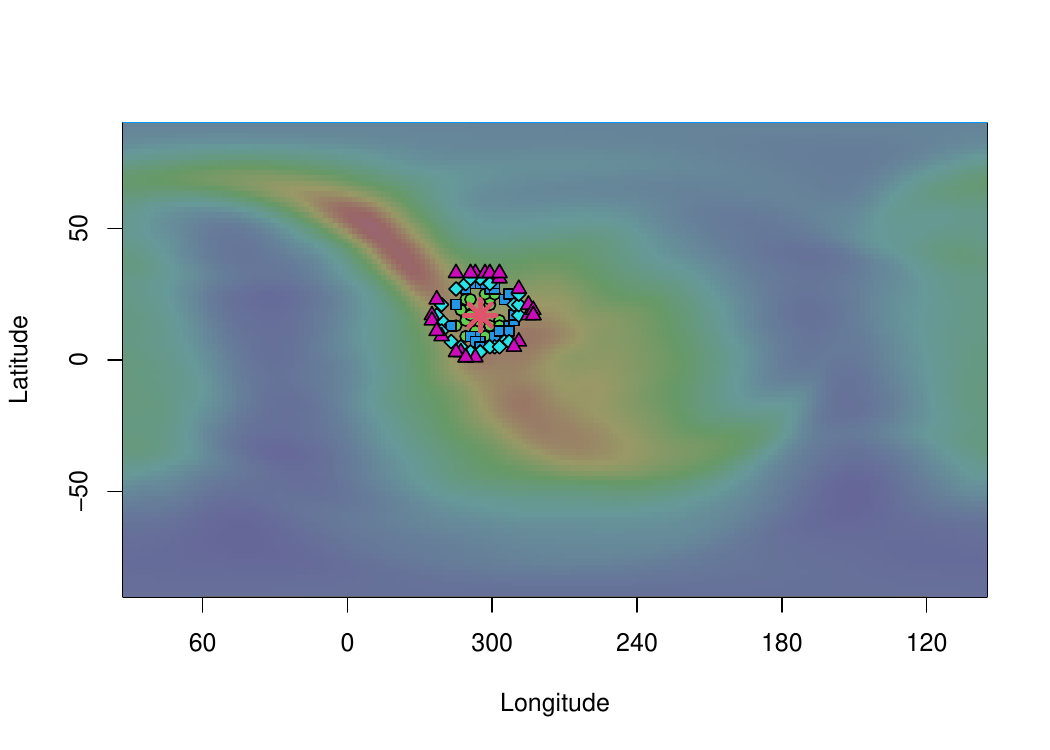}
\includegraphics[scale=0.5, trim=5 0 30 0,clip=TRUE]{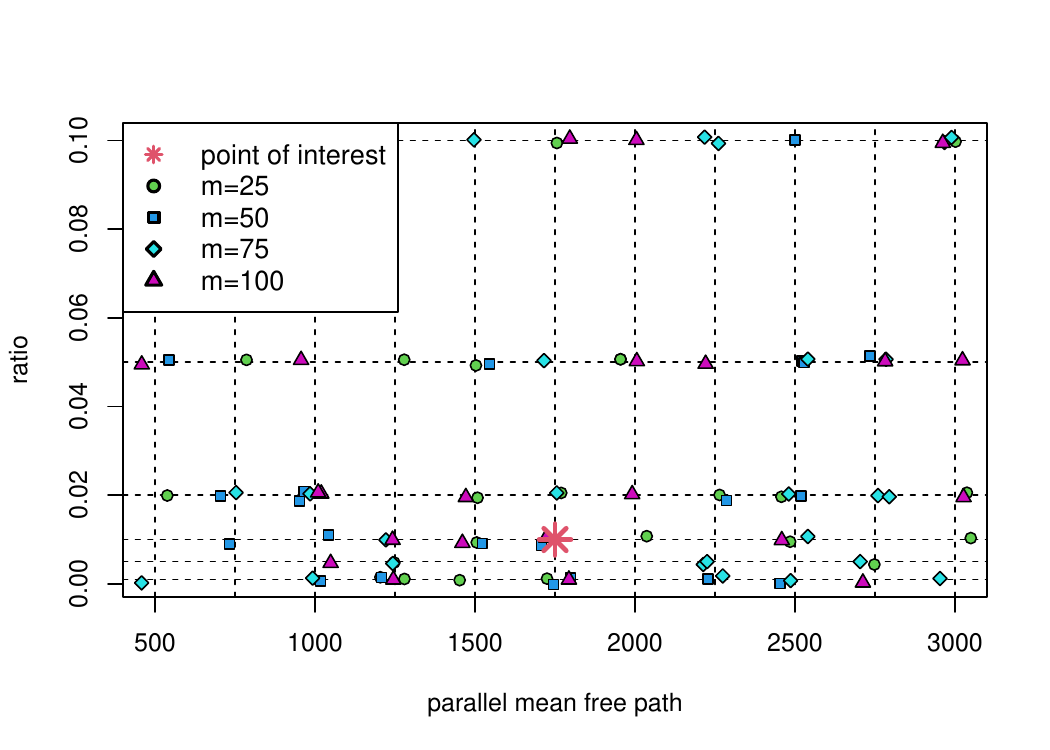}
\caption{Conditioning sets $h(i)$ in a multivariate normal likelihood
(Eq.~\ref{eq:vecchia}) for a single observation $Y_i^M$ using the Scaled Vecchia
approximation. The left panel visualizes neighborhoods in grid space (i.e.\!~$X$
or $X^F$). Model parameter ($u$) space is displayed in the right panel with
jitter added. Intersections of black, dashed lines indicate unique $u$
combinations run on the simulator.
\label{f:ibex_vecchia}}
\end{figure}

Figure \ref{f:ibex_vecchia} illustrates examples of conditioning sets created
by the Scaled Vecchia approximation in the context of IBEX and its associated
simulator. (We will circle back to the inverse problem and the data collected by
the satellite in Section \ref{sec:ibex_data}). Neighborhoods of size $m \in
\{25,50,75,100\}$ are generated for a particular simulator observation $Y_i^M$
(denoted by a red star) from the IBEX computer model. It's difficult to
visualize in higher dimensions, so we break the input space down into separate
2d plots, one for grid space (latitude/longitude) and one for model parameter
space ($u$). But to be clear, points of the same shape and color in both plots
are elements of the same overall conditioning set. For each conditioning set
$h_j(i)$, all points from sets where $m < m_j$ are also included in $h_j(i)$.
That is, green circles from the neighborhood with $m=25$ would also be part of
the conditioning set with blue squares ($m=50$). For reference, we've overlaid
the points in grid space on a grayed-out simulated sky map corresponding with
the observation of interest. Note that more points are concentrated near the
reference point in latitude and longitude, whereas the neighborhood is more
spread out over the unknown model parameters. Therefore, observations condition
on information in close proximity in grid space, but potentially distant in
$u$-space. Furthermore, as $m$ grows, additional points in grid space are mostly
farther away than those already included, but we don't observe the same trend in
$u$-space.

\begin{figure}[ht!]
\centering
\includegraphics[scale=0.5,trim=0 0 0 0,clip=TRUE]{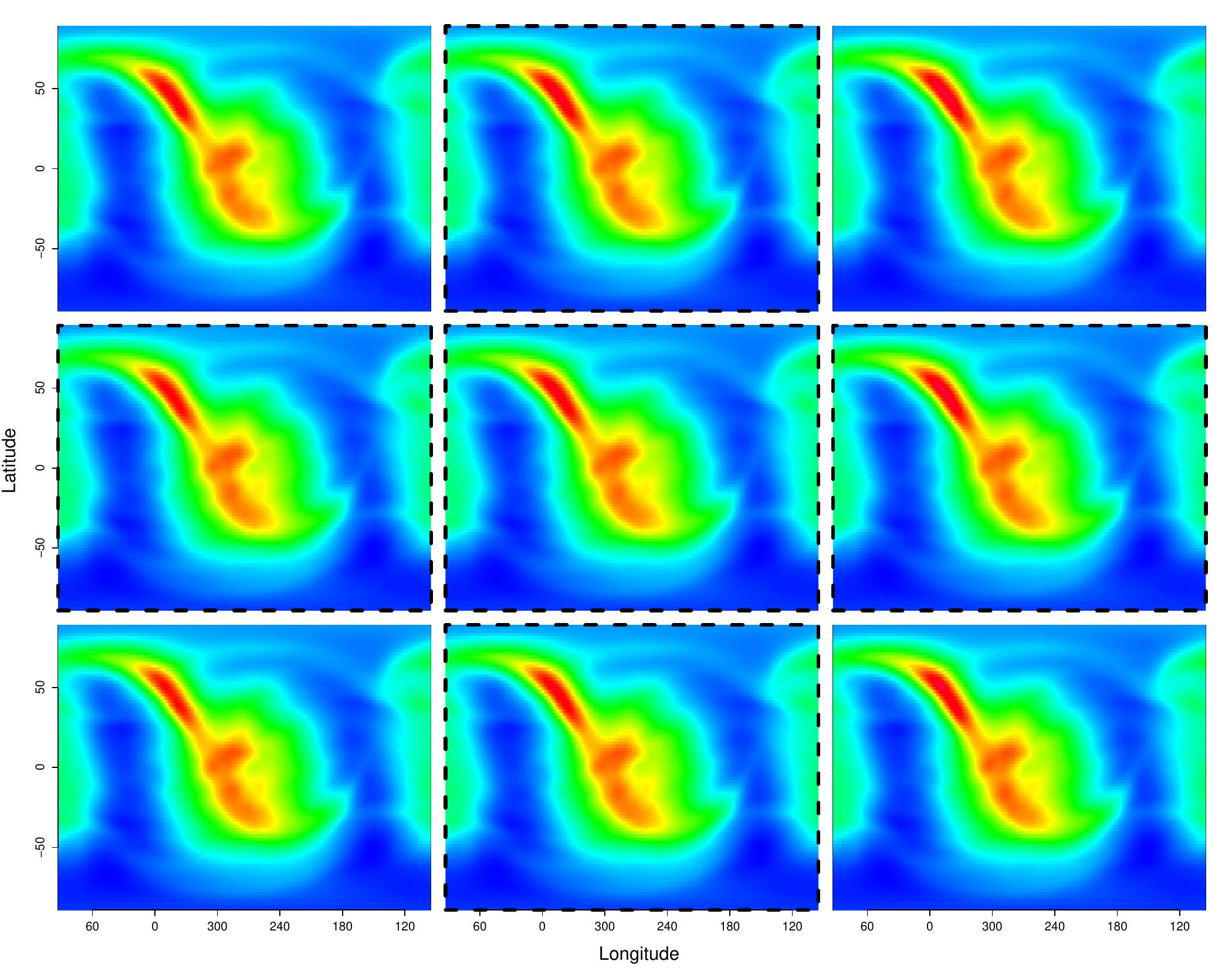}
\caption{Actual and predicted outputs from the IBEX \textit{sky map} computer
model. Corner panels represent the response for four unique combinations of
model inputs. Panels on the cross with bold, dashed borders are predictions at
unobserved input combinations from a GP surrogate fit to $n=66$ runs of the
computer model. Model parameter settings for the predictive panels fall between
the four actual simulator runs shown.
\label{f:ibex_sim_surr}}
\end{figure}

We begin evaluation of Scaled Vecchia's performance on the IBEX simulator with
a visual assessment. Scientists at LANL have provided us with 66 unique model
runs, each producing an image of 16,200 points, or pixels. Figure
\ref{f:ibex_sim_surr} shows four of the actual simulation outputs (corner
panels displayed with thin, solid borders). A Scaled Vecchia surrogate is fit
to all 66 simulated runs. Then, we predict the output for five unobserved
combinations of the model parameters that fall in between the actual runs
shown. These are depicted in the cross of Figure \ref{f:ibex_sim_surr}, those
panels with thicker, dashed borders. To the naked eye, these predictive
surfaces appear to have been generated by the computer model itself. Moving
left to right, or top to bottom, the surfaces change gradually and as one would
expect. Next we show empirically that our surrogate fit is accurate and
outperforms competitors.

\subsection{Empirical Comparison}
\label{sec:emp_comp}

To benchmark our surrogate, we conduct experiments to gauge predictive accuracy
via root mean squared error (RMSE), uncertainty quantification via continuous
ranked probability score \citep[CRPS;][]{gneiting2007strictly}, and
computational efficiency via time. We carry out the same experiment for three
competitors: {\sf R} packages {\sf laGP} \citep{gramacy2014lagp} with defaults
and {\sf deepgp} \citep{deepGP} with 1000 MCMC iterations and {\sf vecchia =
TRUE}, and the current ``status-quo'' SEPIA \citep{gattiker2020lanl} with $n_k
= 3$, implemented in {\sf Python}.

\begin{figure}[ht!]
\centering
\includegraphics[scale=0.45, trim=0 0 0 0,clip=TRUE]{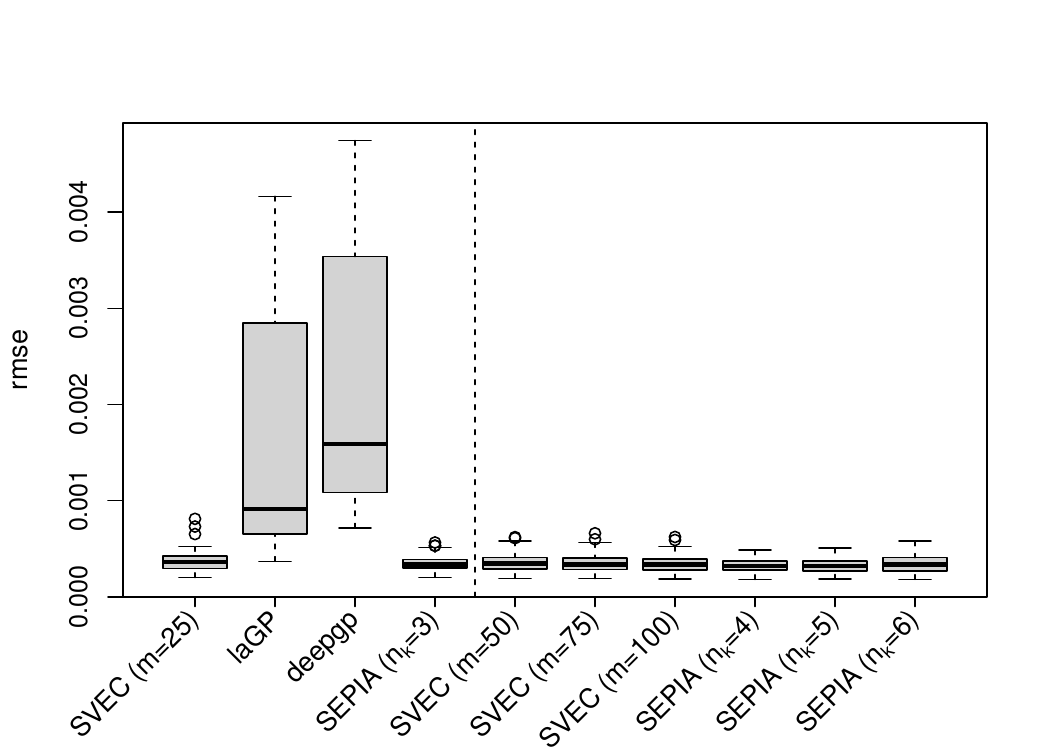}
\includegraphics[scale=0.45, trim=0 0 0 0,clip=TRUE]{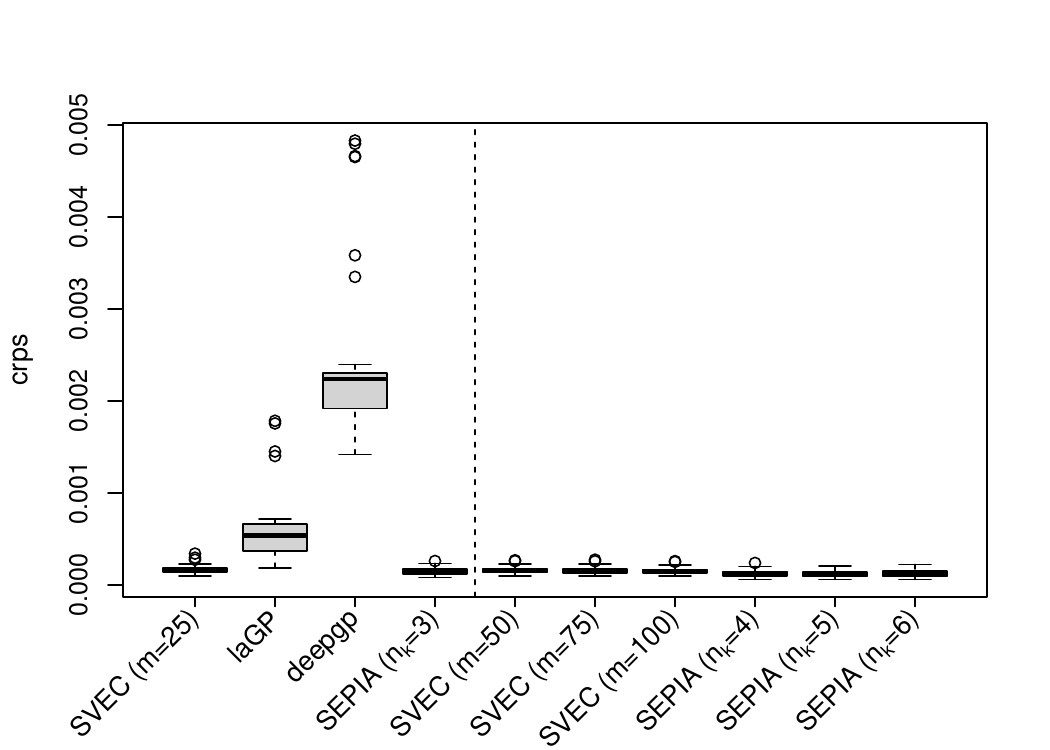}
\caption{Metrics on the IBEX simulator experiment. Smaller is better for both
RMSE and CRPS.
\label{f:ibex_surr_metrics}}
\end{figure}

Figure \ref{f:ibex_surr_metrics} displays the accuracy results of a simple
hold-one-out bakeoff, with time coming next. For each iteration of the
experiment, we fit a Scaled Vecchia GP surrogate on 65 of the $n=66$ model runs
available. Then, we predict at the held out model parameter. For now, just pay
attention to the boxplots on the left-hand side of each panel. We'll come back
to the others in a moment. It is obvious that the Scaled Vecchia approximation
outperforms {\sf laGP} and {\sf deepgp} in both RMSE and CRPS. The local focus
of {\sf laGP} limits its ability to capture global trends, potentially hurting
its performance. Large data and corresponding expensive matrix operations
restrict the number of iterations {\sf deepgp} can run and thus prevents the
MCMC from converging to the target posterior. SEPIA performed similarly to our
Scaled Vecchia approximation, illustrating why it has had such staying power
over the past two decades. As a practitioner, it appears to be a toss-up between
these two methods for which to use. But that's not the entire story.

\begin{figure}[ht!]
\centering
\includegraphics[scale=0.57, trim=0 0 0 0, clip=TRUE]{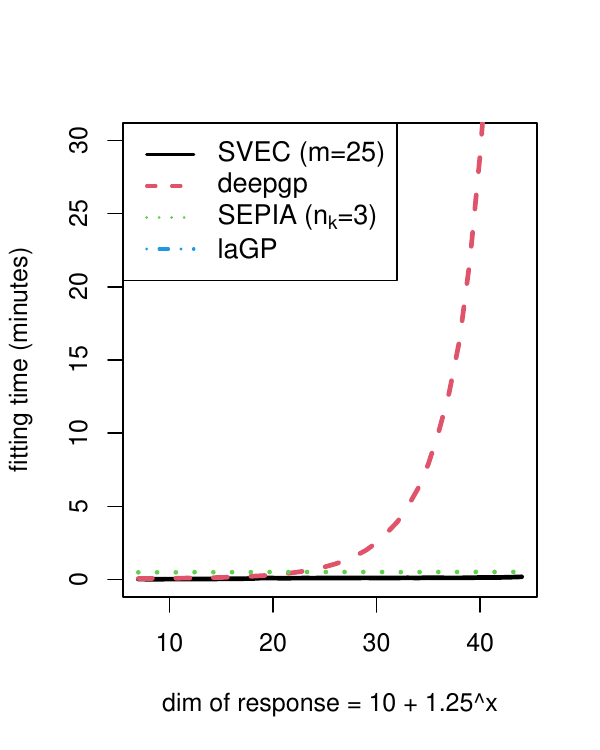}
\includegraphics[scale=0.57, trim=0 0 0 0, clip=TRUE]{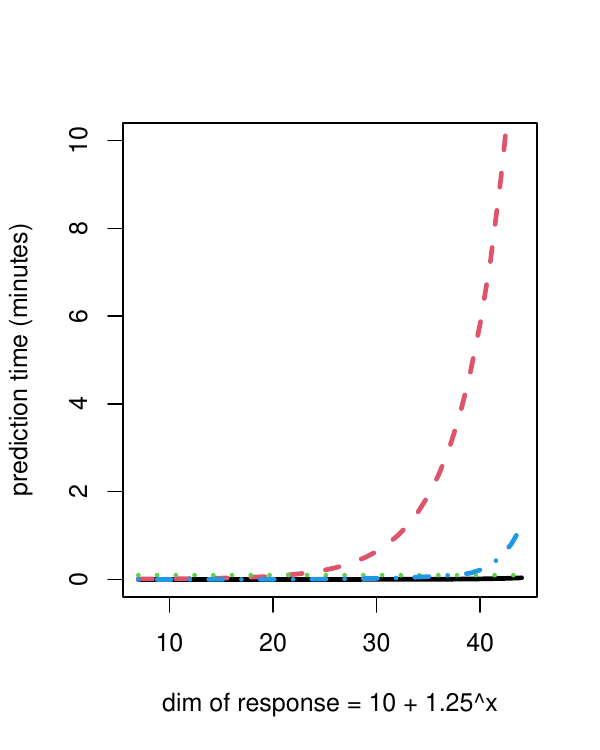}
\includegraphics[scale=0.57, trim=0 0 0 0, clip=TRUE]{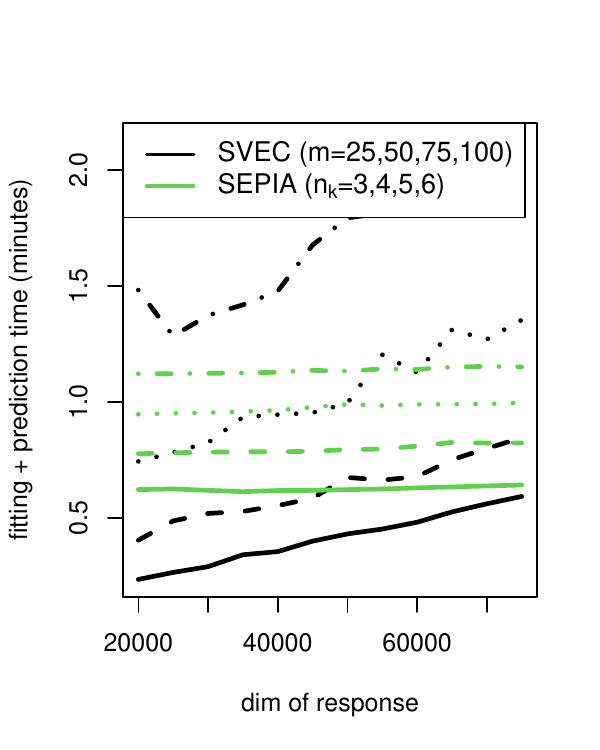}
\caption{Timing metrics on the IBEX simulator experiment. A comparison of the
fastest methods is shown in the right panel, with different line types
indicating neighborhood size ($m$) and number of principal components ($n_k$)
used for Scaled Vecchia and SEPIA, respectively.
\label{f:ibex_surr_timing_dim}}
\end{figure}

\begin{figure}[ht!]
\centering
\includegraphics[scale=0.57, trim=0 0 0 0, clip=TRUE]{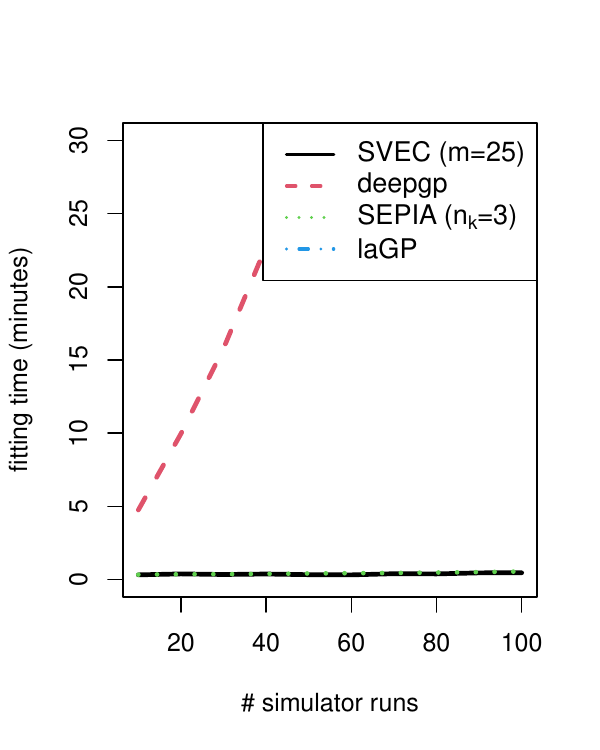}
\includegraphics[scale=0.57, trim=0 0 0 0, clip=TRUE]{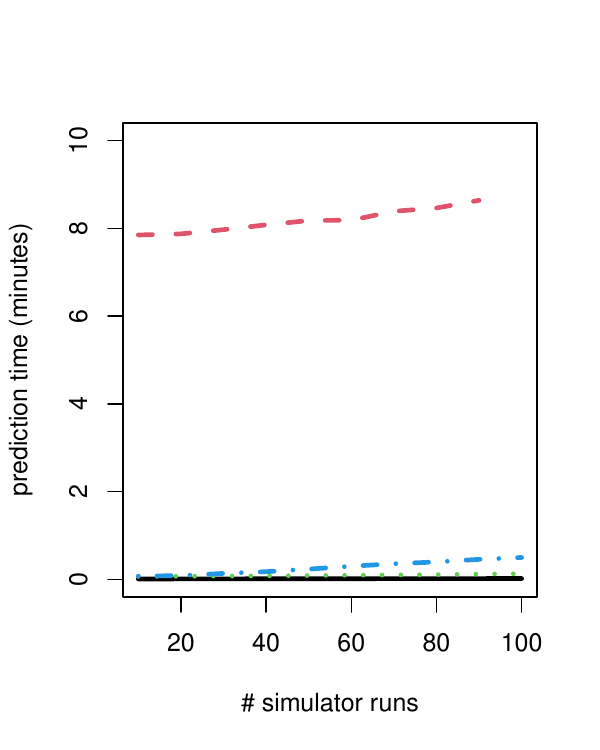}
\includegraphics[scale=0.57, trim=0 0 0 0, clip=TRUE]{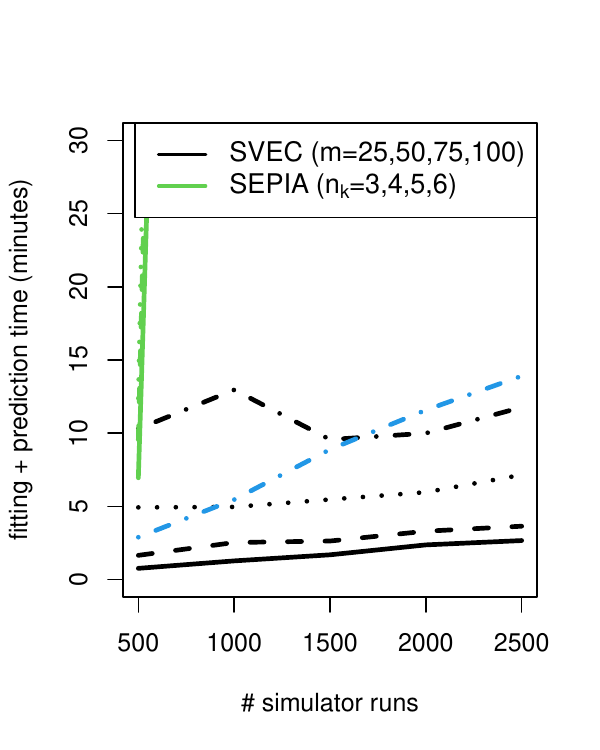}
\caption{Timing metrics on the IBEX simulator experiment.
\label{f:ibex_surr_timing_size}}
\end{figure}

Figures \ref{f:ibex_surr_timing_dim}-\ref{f:ibex_surr_timing_size} summarize
compute times in a similar experiment but varying data size. Two factors exist
that contribute to a computational bottleneck: the dimension of the simulator
response (the main focus of our work), and the number of runs of the computer
model available for training. To test the former, we vary the dimensionality of
the response from 200 to 20,000, while keeping the number of runs constant. For
the two methods that are best equipped for large dimensions (Scaled Vecchia and
SEPIA), we attempt to stretch their capacity by cranking up the length of the
response vector from 20-75K by steps of 5K. At each dimension size, we run five
Monte Carlo iterations for each method and record the average elapsed time.

It is plainly visible from Figure \ref{f:ibex_surr_timing_dim} that Scaled
Vecchia and SEPIA are both well-suited for increasing dimensionality in the
response. Our {\sf deepgp} competitor needs to decompose an $n_M \times n_M$
matrix at each MCMC iteration, which causes it's cost to quickly grow as
dimensionality expands, even aided by the Vecchia approximation. While {\sf
laGP} does better, it is unable to keep pace at a certain point, likely due to
the need to fit a mounting number of local GPs separately for each testing
location. Reducing the neighborhood size for {\sf laGP} could help reduce
computation, but would further degrade it's accuracy. The comparison between
Scaled Vecchia and SEPIA is more nuanced. In the right panel of Figure
\ref{f:ibex_surr_timing_dim}, we see that for a fixed value of $m$, Scaled
Vecchia's runtime increases linearly with expanded dimensionality,
consistent with claims made by \citet{katzfuss2021general}. For a fixed number
of principal components, SEPIA's execution time increases only slightly as
dimension expands, showcasing its utility. However, for response dimension
less than 75K, Scaled Vecchia with $m=25$ still executes faster. And that may
be the largest value of $m$ we need. The boxplots on the right-hand side of
each panel in Figure \ref{f:ibex_surr_metrics} show that increasing $m$ or
$n_k$ doesn't significantly improve performance. It must also be pointed out
that the IBEX simulator surface is not particularly complex over the model
parameters, allowing SEPIA to perform well with a low value of $n_k$. That may
not be the case in other situations, requiring more principal components and by
extension a heavier computational load.

Figure \ref{f:ibex_surr_timing_size} displays timing results from changing the
size of the training dataset (i.e, the number of simulator runs $n$). SEPIA was
specifically built for computer experiments with high-dimensional output. While
it clearly performs well in these situations, it's also fairly inflexible to
other contexts, such as a large number of simulation runs. Scaled Vecchia does
not face this hurdle, as it treats both high-dimensional output and large-scale
univariate datasets similarly, as a scalar response. Our experiment is set up
similar to the previous one, but we keep the dimensionality of the response at
10K while sliding the number of simulator runs from 10 to 100. Again, all
methods are run over five MC iterations for each value of $n$. For the higher
performing methods, we push the campaign size up further ($n=2500$).

It's no surprise that {\sf deepgp} once again lags in run time. Even a modest
sized set of runs takes well over an hour to fit and needs almost 10 minutes to
make a single prediction. Most interesting is the behavior of SEPIA in the
right panel of Figure \ref{f:ibex_surr_timing_size}. For $n > 500$, SEPIA
takes hours to fit and make predictions. In this context, even {\sf laGP}
performs better. Our preferred method, Scaled Vecchia, shows consistency in
scaling linearly with an increased campaign size and executes faster than each
competitor when $m<50$, and for all values of $m$ when $n>1500$.



\section{IBEX Satellite Data}
\label{sec:ibex_data}

We now return to the IBEX inverse problem. We wish to determine the settings $u$
of a computer model that produce a sky map most likely to have generated
observed counts and, more specifically, quantify the uncertainty on our estimates of $u$. Before diving in, we dispense a few important details about
IBEX observations and computer simulations.  We then test our modeling apparatus
on synthetic field observations (computer model at known $u^\star$ and with
Poisson counts) to ensure good performance within the model class, and where we
know the true calibration parameter.  Finally, we report on our calibration of
the actual IBEX satellite observations.

\subsection{IBEX details}
\label{sec:ibex_details}

Each row of the satellite dataset contains Cartesian coordinates ($x$, $y$,
$z$) for the direction the satellite was pointing, time the satellite was aimed
at that location (in seconds), counts of ENAs observed at one of six different
energy levels (ESA), and a background rate $\lambda_b$ (ENAs/sec). Note that
for the sake of brevity, we focus only on ENAs collected at energy level ESA 4.
More important to consider is that ENAs are known to come from sources other
than the heliosphere. Space scientists quantify this noise for each location
($X^F$) viewed by the satellite and add it as a background rate ($\lambda_b$)
that must be accounted for in the data. For example, the left panel of Figure
\ref{f:fig1} displays ENA rates {\em after} $\lambda_b$ has been subtracted
out. Although this rate is an estimate rather than an exact value, we follow
the lead of other researchers \citep{osthustheseus2023} and treat it as
constant in our work. For the purpose of solving the inverse problem, we
address this through an update to the mean expression in the Poisson likelihood
in Eq.~\ref{eq:pois_inv_bayes_mod} such that $\lambda(X^F) = m(u, X^F) -
\lambda_{b}(X^F)$.

We have referred to the IBEX simulator throughout this paper without going into
much detail. While we leave a more in-depth physical interpretation and
analysis of parameters to more appropriate venues
\citep{zirnstein2021heliosheath,zirnstein2025global,swaczyna2020density,huang2025numerical},
some basic context is provided here. We focus on two particular models
available (among many): one for the GDF and one for the ribbon. In the ribbon
model, the parameter \textit{parallel mean free path} is the distance a
hydrogen ion travels beyond the heliopause, or edge of the heliosphere, before
interacting with another particle, receiving an electron and becoming an ENA,
and ``turning back.'' The \textit{ratio} parameter is the fraction of
perpendicular mean free path, another measure of distance, and \textit{parallel
mean free path}. These are our $u$ values. The response from the GDF model is
static and is merely added to the output of the ribbon simulator. This is not
the case for GDF simulators in general, but the GDF model we have access to
does not vary with any parameters of interest.

Scientists know that the makeup of the heliosphere evolves over time, largely
due to variations in the solar wind that accompany the approximately 11 year
solar cycle. Physicists build these changes into their simulators either through
additional parameters, restricting the application of their models to a specific
timeframe, or simply averaging over the solar cycle. The GDF model we utilize
was developed to correspond to the years 2009-2011, while the ribbon model
averages the composition of the heliosphere over the entire solar cycle. We
factor in these heliospheric variations by only comparing satellite data from
the years 2009-2011 to simulator runs.

\subsection{Synthetic testbed for inverse problem validation}
\label{sec:synth_data}

\begin{figure}[ht!]
\centering
\includegraphics[scale=0.55,trim=5 15 95 45,clip=TRUE]{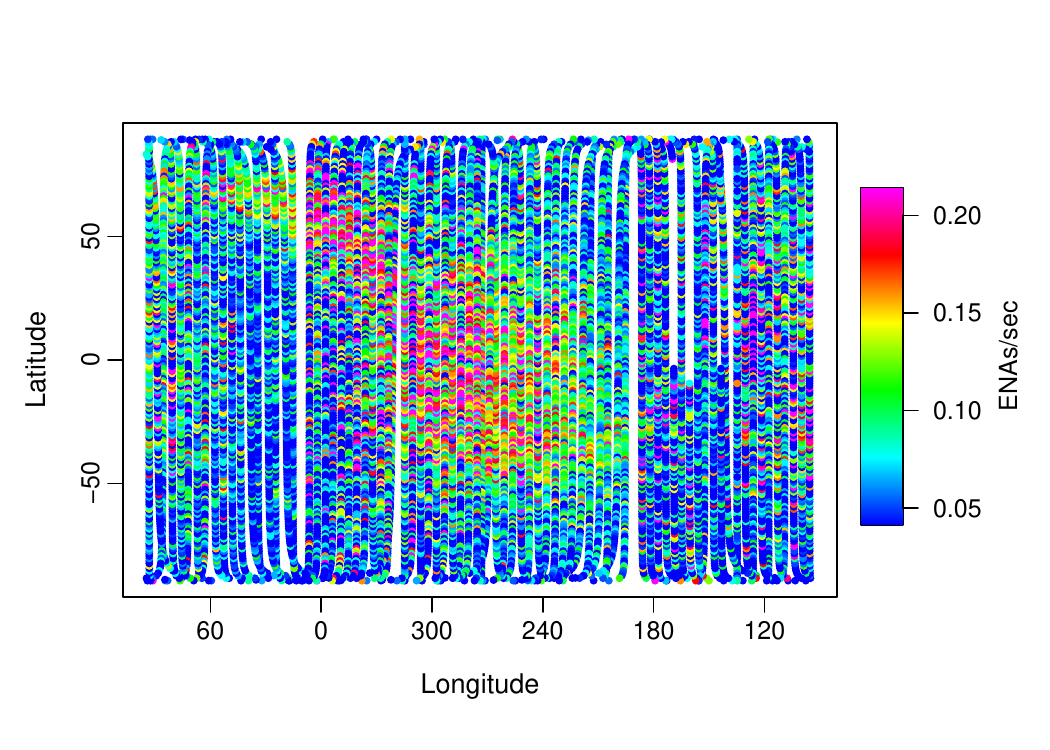}
\includegraphics[scale=0.55,trim=5 15 15 45,clip=TRUE]{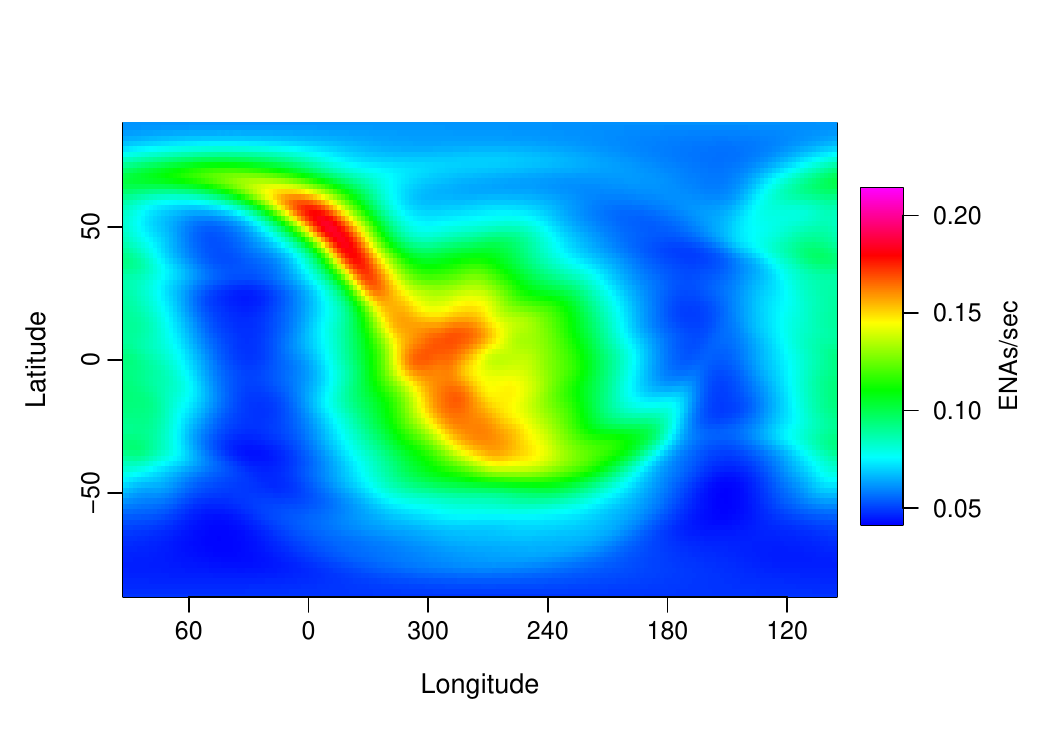}
\includegraphics[scale=0.55,trim=5 15 95 45,clip=TRUE]{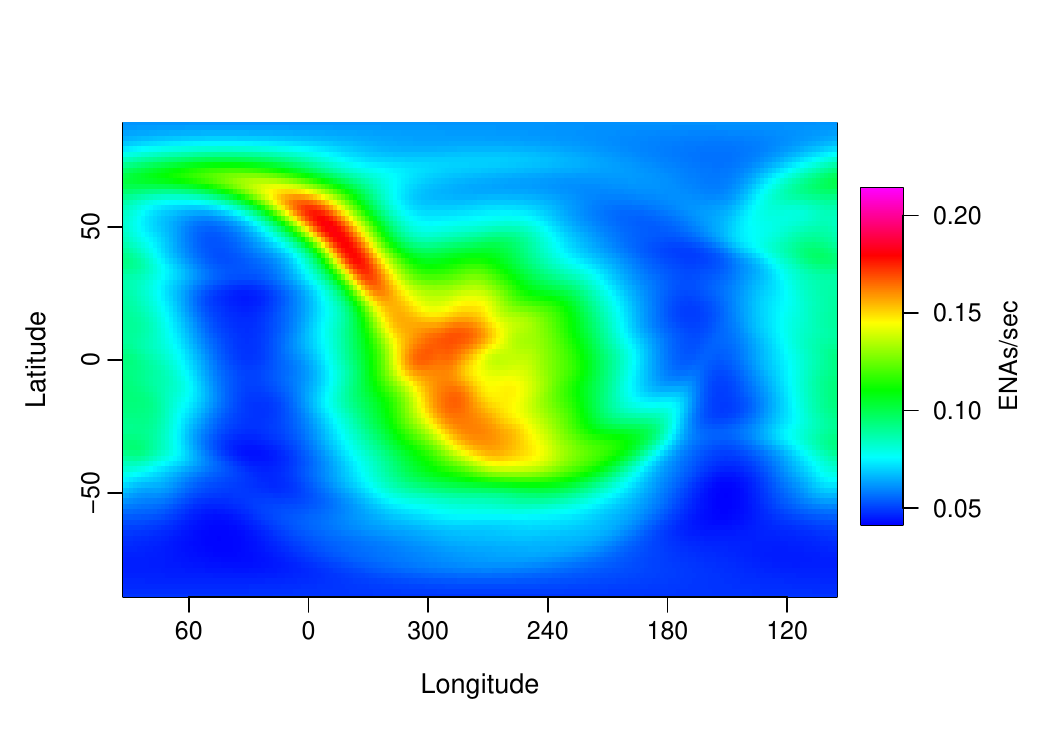}
\includegraphics[scale=0.55,trim=5 15 15 45,clip=TRUE]{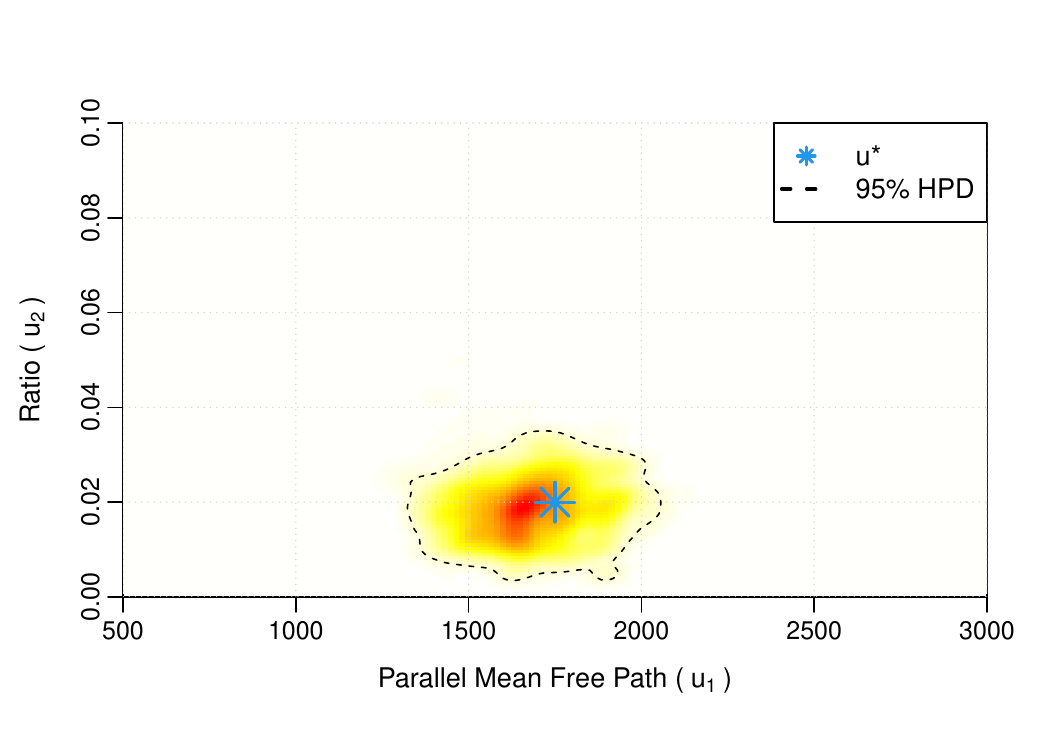}
\caption{Estimate of ENA rate
(counts/seconds) based on synthetic Poisson draws from a ``true'' computer
model setting and varying exposure times (top left). IBEX simulator output of
ENA rate at \textit{parallel mean free path} $= 1750$ and \textit{ratio} $=
0.02$ (top right). Predicted output via a fitted Scaled Vecchia surrogate at
estimated calibration parameters (bottom left). Posterior samples of model
parameter $u$ as a kernel density (bottom right).
\label{f:ibex_synth_ex}}
\end{figure}

Using the corpus of $n=66$ computer model runs, varying a design of
\textit{parallel mean free path} and \textit{ratio} $u$-values, described in
Section \ref{sec:vecchia}, we select a value $u^\star$ and set the corresponding
output ($\lambda^\star = m(u^\star, X^M)$) as the ``true'' underlying mean ENA
rate (upper right panel of Figure \ref{f:ibex_synth_ex}). Next, we gather
measurement locations $X^F$, exposure times $e$, and background rates
$\lambda_b$ from a single year that the IBEX satellite has been in operation,
ignoring the observed counts of ENAs in this fabricated example. For each
latitude/longitude pair (i.e.~a row in $X^F$), we draw synthetic satellite (or
field) data from a Poisson, $Y_i^F \sim
\mathrm{Pois}(\lambda^\star_i + \lambda_{b_{i}}, e_i)$, where $\lambda_i^\star$ is determined by the
computer model $m(\cdot,\cdot)$. The top left panel of Figure
\ref{f:ibex_synth_ex} shows the result, which we will regard as data we
``observed.'' Note that observations are not on a fine grid, but instead match
the recorded coordinates for ENAs captured by the IBEX satellite in its orbit.
This accounts for the unique spread of points and the strings of missing data,
which correspond to certain orbits.

We run Algorithm \ref{alg:gibbs}, treating this synthetic data as $(X^F, Y^F)$
and fitting a Scaled Vecchia GP surrogate on the provided computer model output
(omitting responses for $u^\star$) as $(X^M, \lambda^M)$. It should be noted
that we employ an empirical Bayes approach
\citep{berger1985statistical,kennedyohagan2001}, pre-estimating the GP
hyperparameter $\theta$ and only fitting the surrogate on simulator output
\citep{Liu2009ModularizationIB}. We execute 10,000 MCMC iterations, throwing
away 1,000 as burnin and thinning by 10. Figure \ref{f:ibex_synth_ex} displays
the remaining posterior samples of $u$ (bottom right) as a kernel density along
with a 95\% highest posterior density (HPD) interval. The posterior mean of $u
\mid Y^F$ lies near $u^\star$, demonstrating that our framework is able to
concentrate mass around the ``true'' model parameter. As a sanity check, we
take the estimated posterior mean and plug it in to our fitted Scaled Vecchia
GP surrogate of the IBEX simulator. The predicted output is shown in the bottom
left panel of Figure \ref{f:ibex_synth_ex}. There are some nearly imperceptible
differences between the estimated sky map and the true $\lambda^\star$ (i.e,
the shape and intensity of the ribbon vary slightly between the two). However,
from a purely visual standpoint, it appears that each map could have reasonably
generated the data we ``observe'' in the top left panel.

\begin{figure}[ht!]
\centering
\includegraphics[scale=1.0,trim=0 0 0 0,clip=TRUE]{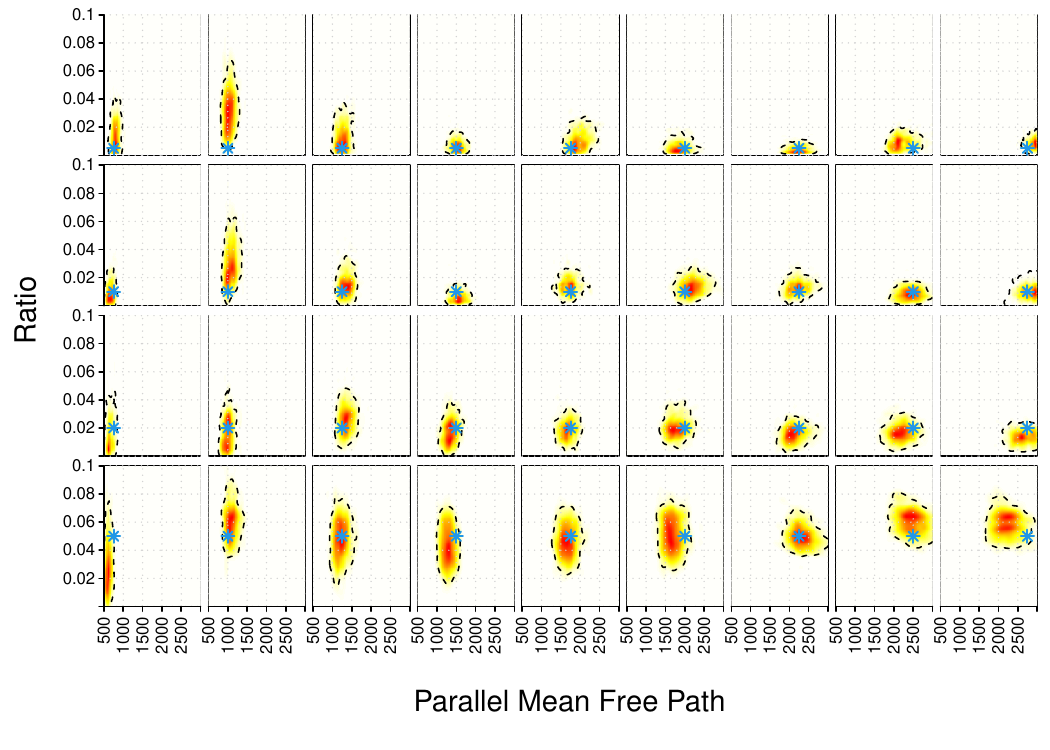}
\caption{Posterior distributions $u \mid Y^F$ and 95\% HPD intervals from runs
of our Poisson Bayesian inverse problem framework on 36 different synthetic
datasets. Blue stars indicate the value $u^\star$ from which the data was
simulated.
\label{f:synth_estimates_all}}
\end{figure}

To further validate our methodology, we apply this method to each unique model
parameter available. We have access to simulator runs representing all
combinations for 11 unique values of parallel mean free path and 6 unique values
of ratio ($n=66$). The spread of parameter values is determined by space
scientists to include all values considered reasonable: parallel mean free path
ranges from 500-3000 astronomical units (AU); ratio values are unitless, and
span from 0.001 to 0.1. Due to the inability for the model to furnish posterior
samples beyond the boundary parameters, we do not test those combinations for
which at least one parameter is on its edge (i.e, when parallel mean free path
$\in \{500, 3000\}$ and/or ratio $\in \{0.001, 0.1\}$). Removing these edge
cases results in 36 remaining test runs. For each combination, we follow the
same procedure as above: select one model parameter and set as $u^\star$, create
simulated observed satellite data, fit a  Scaled Vecchia surrogate on training
data excluding $u^\star$, and run Algorithm \ref{alg:gibbs} to sample from the
posterior $u \mid Y^F$. Results are shown in Figure \ref{f:synth_estimates_all}.

Each panel shows posterior samples of $u \mid Y^F$ for a unique $u^\star$ as a
kernel density. We have indicated each $u^\star$ that generated synthetic $Y^F$
as a blue star. Our 95\% HPD intervals encompass the truth in 34 of the 36 of
the iterations, near the nominal coverage rate we'd expect. Therefore, we can
confidently assert that our proposed framework is able to recover the ``true''
parameter values from \textit{synthetic} satellite data. Next, we apply this to
real data, with the goal of helping scientists better understand the behavior of
the heliosphere by estimating these parameters for actual ENA counts collected
by the IBEX satellite.

\subsection{Real IBEX ENAs}
\label{sec:ibex_exp}

\begin{figure}[ht!]
\centering
\includegraphics[scale=0.41,trim=10 15 100 45,clip=TRUE]{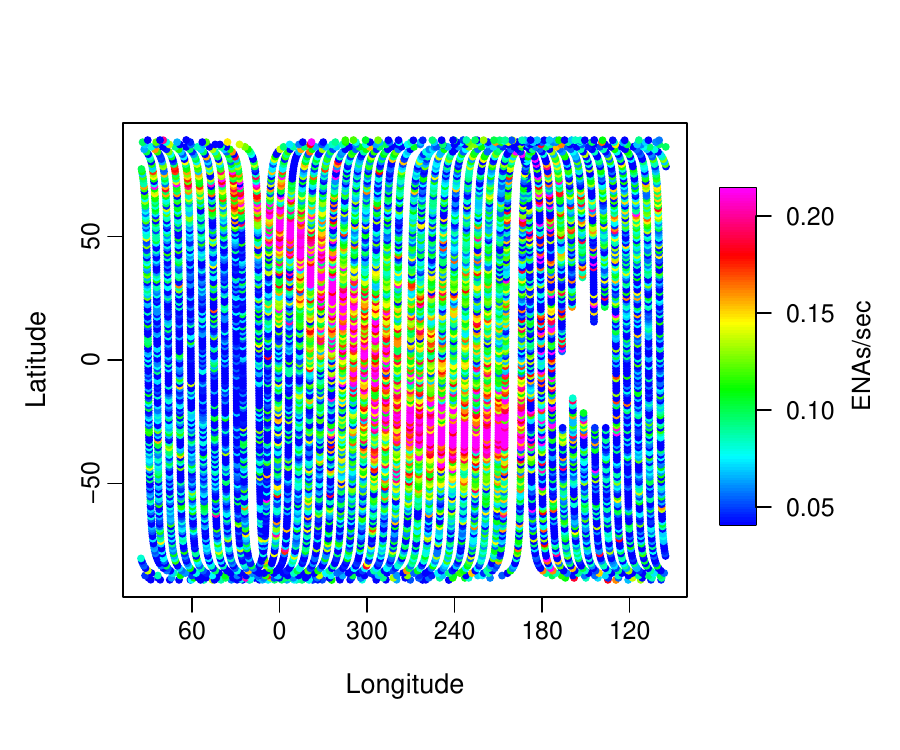}
\includegraphics[scale=0.41,trim=55 15 10 45,clip=TRUE]{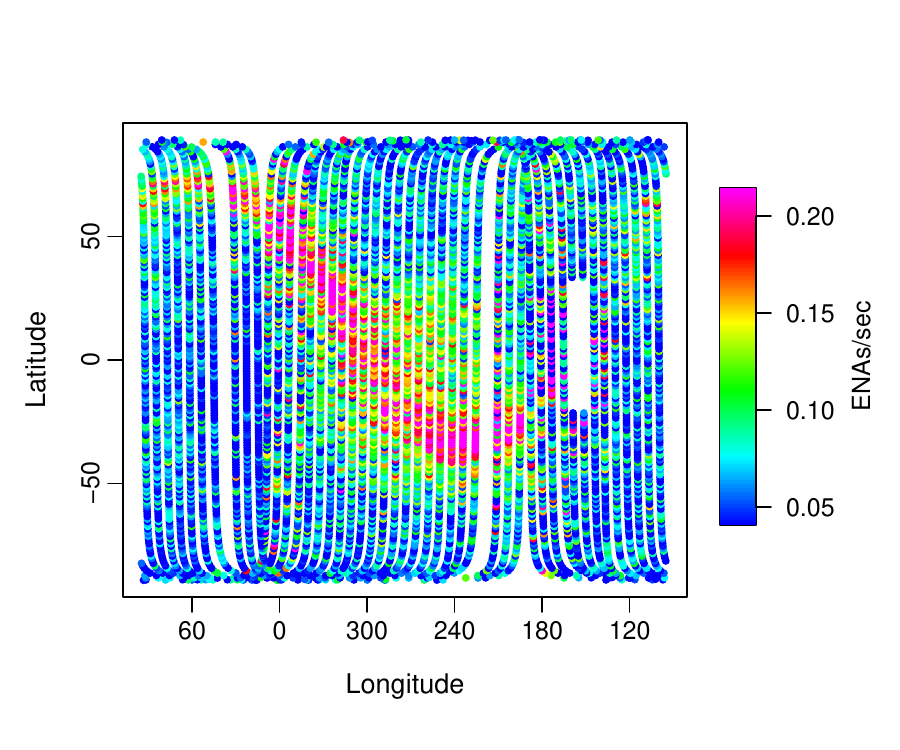}
\includegraphics[scale=0.41,trim=0 15 15 45,clip=TRUE]{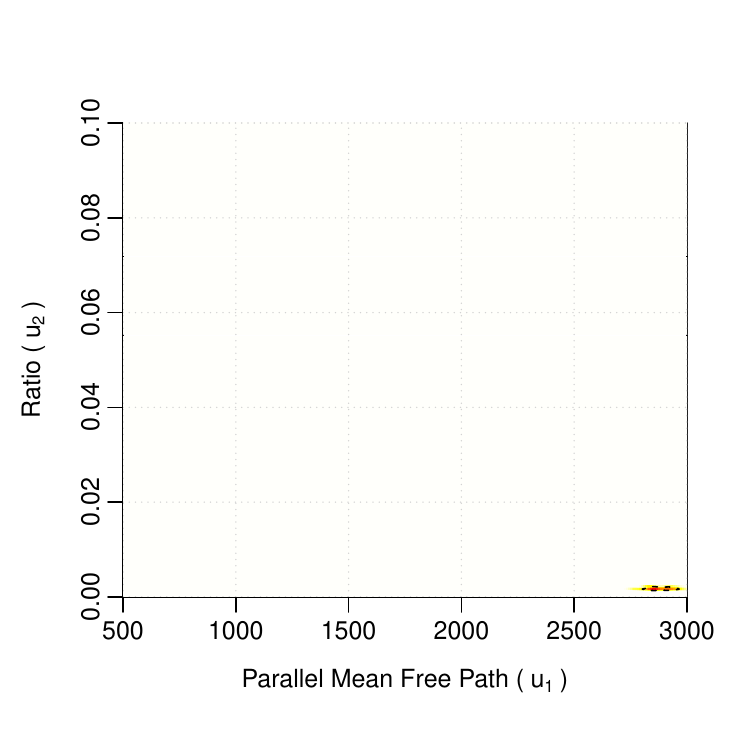}
\includegraphics[scale=0.41,trim=10 15 100 45,clip=TRUE]{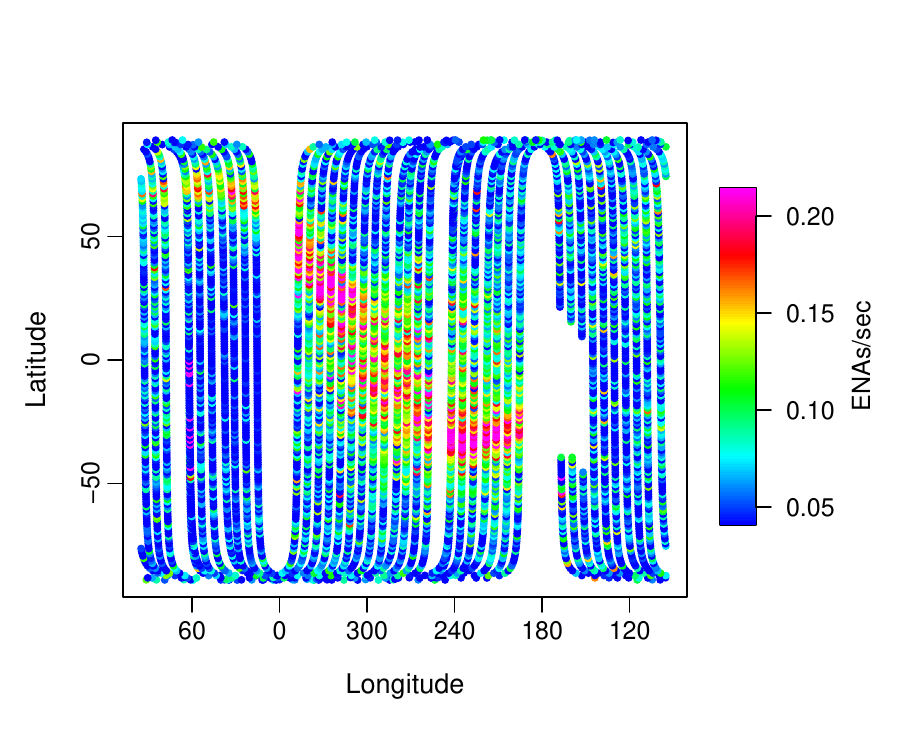}
\includegraphics[scale=0.41,trim=55 15 10 45,clip=TRUE]{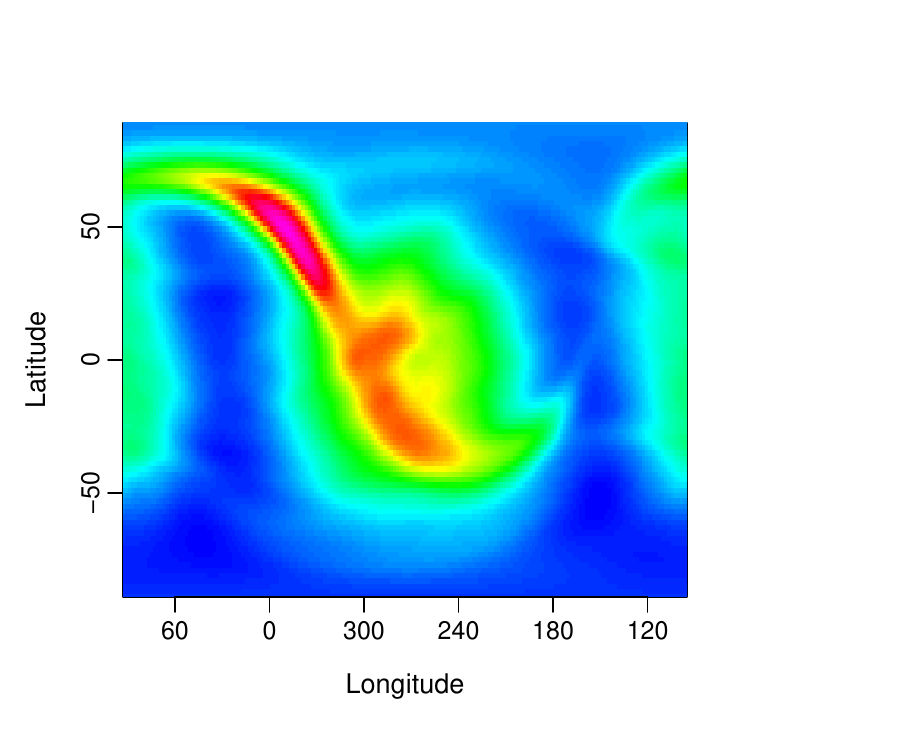}
\includegraphics[scale=0.41,trim=0 15 15 45,clip=TRUE]{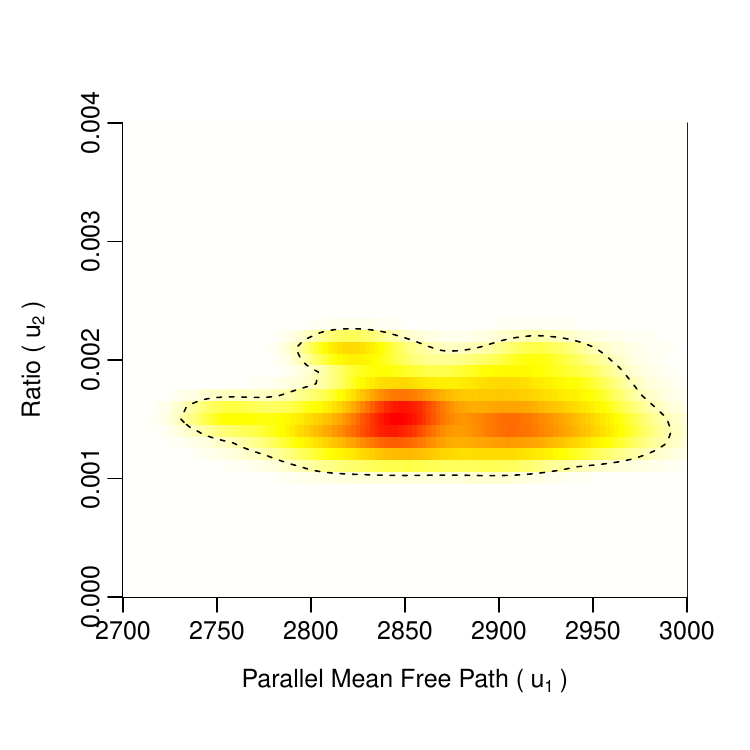}
\caption{Observed IBEX satellite data for years 2009-2011 (top left, top middle, bottom left). Predicted output via a fitted Scaled Vecchia surrogate at
estimated model parameters (bottom middle). Two views for the posterior
distribution of sampled model parameters $u$ (top right, bottom right).
\label{f:ibex_real_ex}}
\end{figure}

Our experimental setup will be similar to that in Section \ref{sec:synth_data},
only substituting in real data for simulated data. Space scientists at LANL have
provided us with satellite data from IBEX for the years 2009-2011, aligning with
the implementation of the computer model. The number of records in each sky map
varies across the three years from $\sim$7-9K observations, giving us over 23K
rows of field data ($X^F, Y^F$) for use in our analysis. We execute Algorithm
\ref{alg:gibbs} by fitting a Scaled Vecchia GP surrogate to \textit{all} 66
computer model runs and sample from the posterior $u \mid Y^F$ via MCMC. One
difference from Section \ref{sec:synth_data} is that field data ($X^F, Y^F$) are
not exclusively from one map (Figure \ref{f:ibex_real_ex}). Instead, ($X^F,
Y^F$) is formed by stacking the counts, exposure times, background rates, and
Cartesian coordinates $x$, $y$, $z$ from years 2009-2011.

Figure \ref{f:ibex_real_ex} makes it evident that the predicted GP surrogate
output at the posterior mean of $u \mid Y^F$ marks a clear difference from the
trends observed in real data. In particular, the intensity of the ribbon in the
surrogate prediction near latitude $-30^\circ$ and longitude $240^\circ$ appears
to be much lower than what is observed in the satellite data for each year.
There's also a gap (in this instance overestimating the ENA rate) between
surrogate and reality between longitudes $60^\circ$ and $120^\circ$ and
latitudes $\pm50^\circ$. Another indication of this clash is visible in the
right panels of Figure \ref{f:ibex_real_ex}. Note that these both display the
same posterior distribution, but differ in their axis limits. Posterior mass is
concentrated in the extreme corner of the model parameter space. Our priors
attempt to push sampling away from the boundaries, but the likelihood dominates
the posterior. We take this as an indication that the computer model is lacking
in its attempt to represent reality, and the Metropolis sampler seeks to go
beyond the range of model parameters present in the training data.

The boundary-pushing estimate for $u$ and the gap, which is plainly visual
between surrogate predictions and collected data, are somewhat unsatistfying. We
address in more detail the potential discrepancy between simulator and reality
in Section \ref{sec:discuss}. But we wish to offer some validation that our
method performs well for the real data, \textit{given} our provided computer
model implementation and the available runs. To do so, we ran a final experiment
using cross validation and calculation of CRPS at testing locations. We split
the satellite data from 2009-2011 into ten equally-sized folds. For each fold
$f$, we ran our Poisson Bayesian inverse problem framework as before, but only
use the other nine folds as field data, and store the resulting model parameter
estimate, $\hat{u}_f$. Note that for each fold, all computer model output is
still used to train the surrogate. We then predict the underlying ENA rate
$\lambda$ at the held out field data locations in three separate manners: 1)
over a fine grid of parallel mean free path, with ratio fixed at the posterior
mean of $u_f$; 2) over a fine grid of ratio, with parallel mean free path fixed
at the posterior mean; and 3) over a fine grid of both parallel mean free path
and ratio. We calculate CRPS for all predictions.

\begin{figure}[ht!]
\centering
\includegraphics[scale=0.53,trim=16 0 26 0,clip=TRUE]{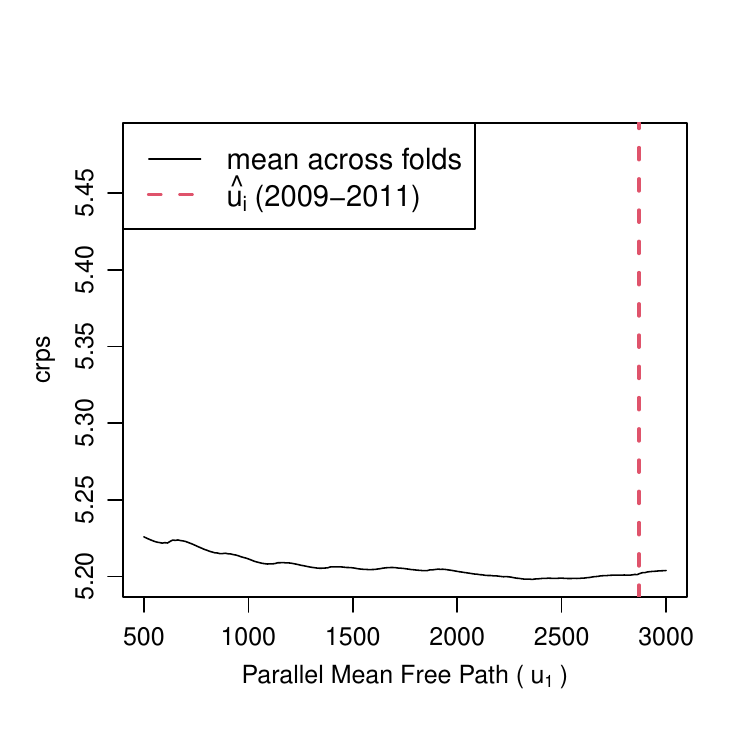}
\includegraphics[scale=0.53,trim=54 0 28 0,clip=TRUE]{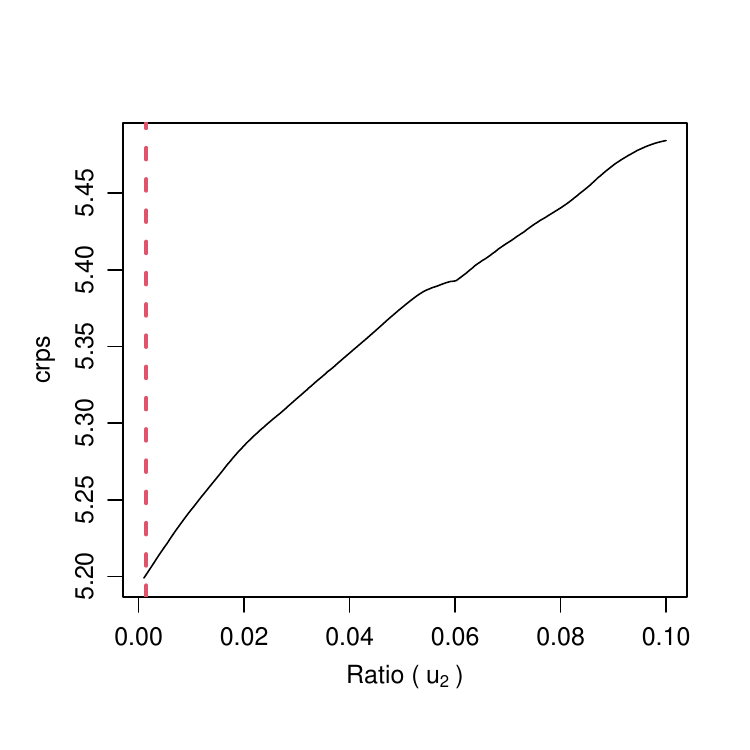}
\includegraphics[scale=0.53,trim=12 0 22 0,clip=TRUE]{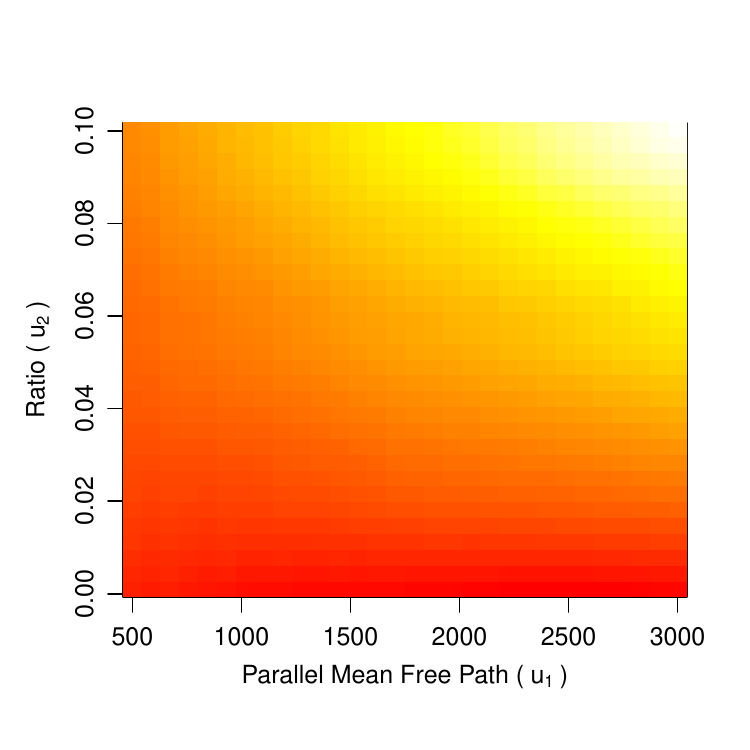}
\caption{Metrics from a cross-validation experiment for IBEX satellite data.
Averaged CRPS across folds on fine grids of parallel mean free path and ratio
(left and middle) and a lattice over both (right). In the right panel, low
values are in red and high values are in yellow. Lower CRPS indicates better
performance.
\label{f:real_cv_valid}}
\end{figure}

The outcome of this experiment is summarized in Figure \ref{f:real_cv_valid}.
The left and middle panels display CRPS for surrogate predictions using 200
equally-spaced values of parallel mean free path and ratio, respectively, at
held-out satellite data locations. CRPS is averaged over the ten folds. For
reference, we include a red dashed line to mark our estimate of the model
parameter $u_i$ when using all the IBEX data from 2009-2011. It's clear that
the setting of the ratio parameter largely determines how well the surrogate
predictions fit the real data. CRPS changes significantly across the entire
range of ratio, but only shifts slightly for parallel mean free path.
Furthermore, looking back at Figure \ref{f:ibex_real_ex}, note that the
posterior (bottom right panel) is spread across a broad set of values for
parallel mean free path ($\sim$2700-3000). That ridge or plateau-like behavior
lines up with our results here, as that spectrum of parallel mean free path
corresponds to ones that lead to low values of CRPS (left panel) in this
cross-validation experiment. The right panel of Figure \ref{f:real_cv_valid}
shows CRPS, averaged across ten folds, for surrogate predictions on a
$30\times30$ grid over both model parameters. There appears to be some
interaction between the two parameters. Adjusting parallel mean free path when
ratio $= 0.1$ greatly affects CRPS, going from red-orange to yellow-white.
However, when ratio $= 0.001$, varying parallel mean free path stays fairly
constant. The importance of parallel mean free path therefore lessens as ratio
decreases in value, consistent with the posterior in Figure
\ref{f:ibex_real_ex}.

The take-home message from Figure \ref{f:real_cv_valid} is that the mean of
posterior samples $u \mid Y^F$, when using the complete satellite dataset and
the \textit{available} computer model output, performs well in terms of CRPS.
That is, our estimate of $u$ resides in the region with settings of parallel
mean free path and ratio that result in the lowest values of CRPS across the
supplied parameter bounds. Therefore, given the constraints (i.e, bias from
reality) of a specific computer model and the $u$-values provided in a campaign
of simulator runs, we can be confident that our Bayesian framework concentrates
posterior mass at the best possible value of the model parameter $u$.



\section{Discussion}
\label{sec:discuss}

We have introduced an effective tool for solving inverse problems using Bayesian
methods where 1) the simulator representing the physical model or process is
expensive, limiting the number of runs available for analysis and demanding a GP
surrogate, but 2) the computer model output is extremely high-dimensional,
requiring novel methods to induce sparsity and approximate the GP, and 3) data
from a field experiment are Poisson counts, bucking the current Bayesian inverse
problem literature that focuses on normally distributed field observations. Our
proposed Poisson Bayesian inverse problem framework fills this gap by
introducing a modular approach that satisfies all three requirements. We have
shown that our framework is able to recover the model parameter $u^\star$ at the
nominal 95\% rate when provided with synthetic data where the ``truth'' is known
and all modeling assumptions have been satisfied. We applied this methodology to
a subset of data collected by the IBEX satellite from 2009-2011 and
corresponding simulated heliospheric sky maps from a sophisticated computer
model.

\begin{figure}[ht!]
\centering
\includegraphics[scale=1.0,trim=0 0 0 0,clip=TRUE]{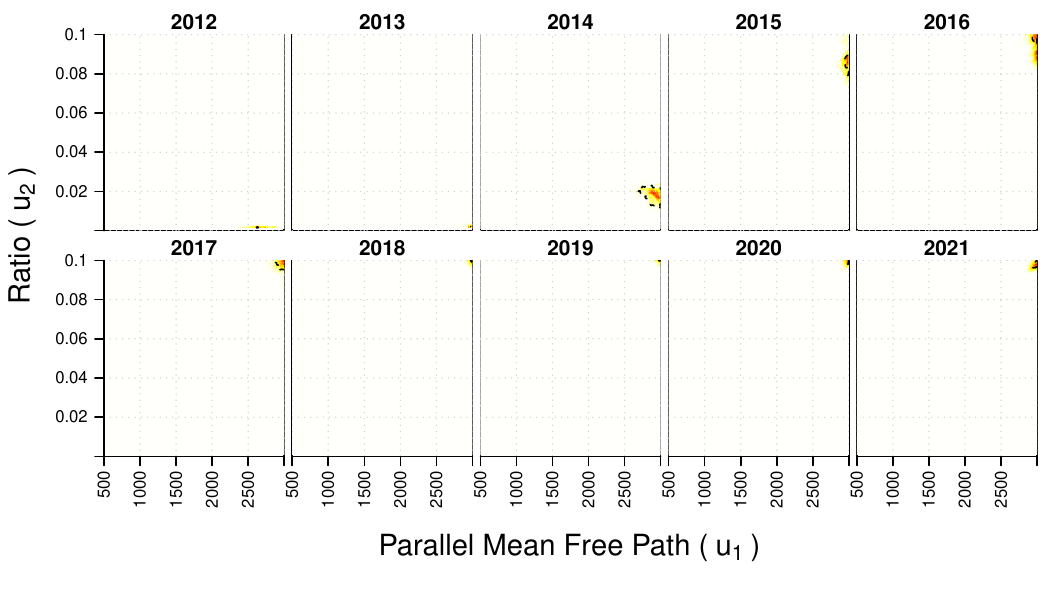}
\caption{Posterior distributions $u \mid Y^F$ and 95\% HPD intervals for 10
separate runs of our Poisson Bayesian inverse problem framework. Field data
$(X^F, Y^F)$ for each run come from a single year between 2012-2021.
\label{f:real_estimates_all}}
\end{figure}

The IBEX satellite has collected counts of ENAs for over 15 years. Our work
focused on data collected in a three-year period for which the computer model we
relied on was aligned. Figure \ref{f:real_estimates_all} shows resulting
estimated posterior distributions for model parameter $u$ when field data is
split by year (2012-2021). Behavior is similar to what we observed in Figure
\ref{f:ibex_real_ex}, but even more extreme, as the Metroplis sampler pushes up
against the boundary of the current range of $u$. We certainly do not claim
these posteriors to be the true distribution of parallel mean free path and
ratio. Instead, we suggest our work be extended to account for un-modeled
discrepancy and paired with computer simulations that integrate other phenomena
not represented in this paper and/or match years outside of 2009-2011.

\begin{figure}[ht!]
\centering
\includegraphics[scale=0.51,trim=0 5 75 0,clip=TRUE]{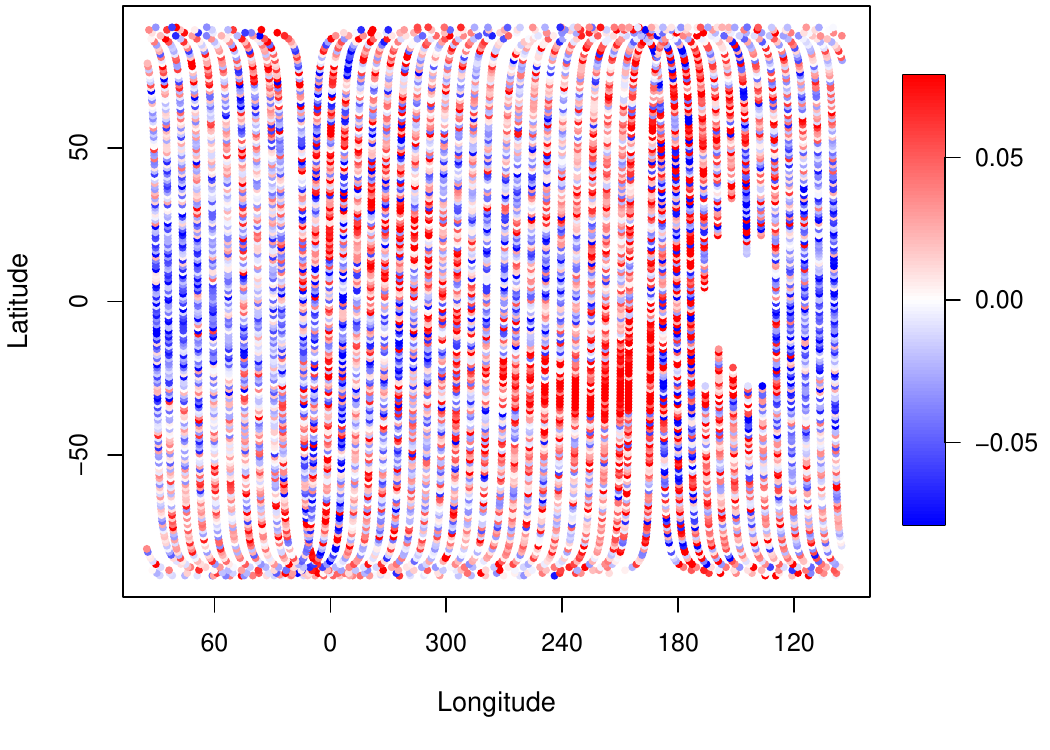}
\includegraphics[scale=0.51,trim=0 5 0 0,clip=TRUE]{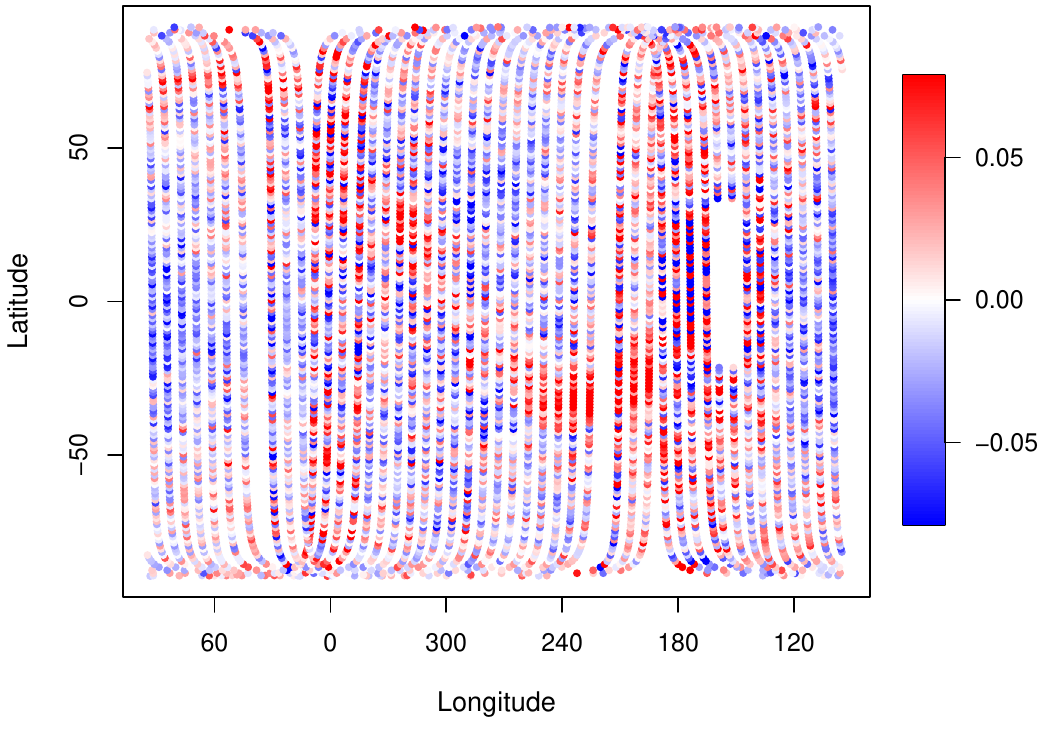}
\includegraphics[scale=0.51,trim=0 5 75 0,clip=TRUE]{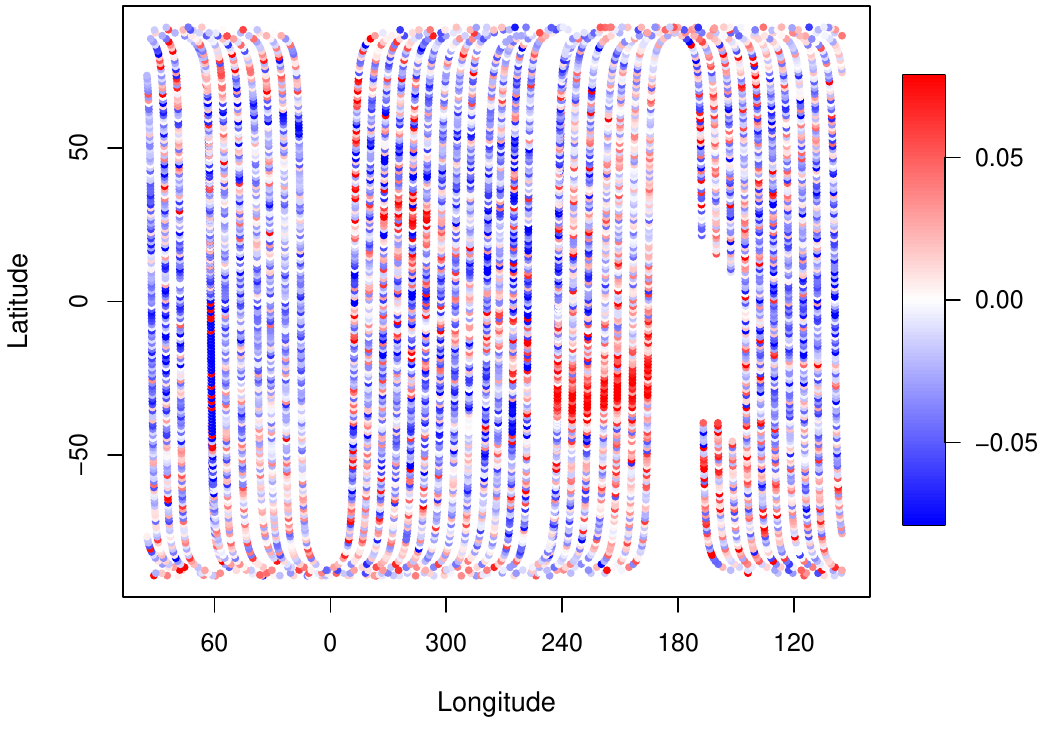}
\includegraphics[scale=0.51,trim=0 5 0 0,clip=TRUE]{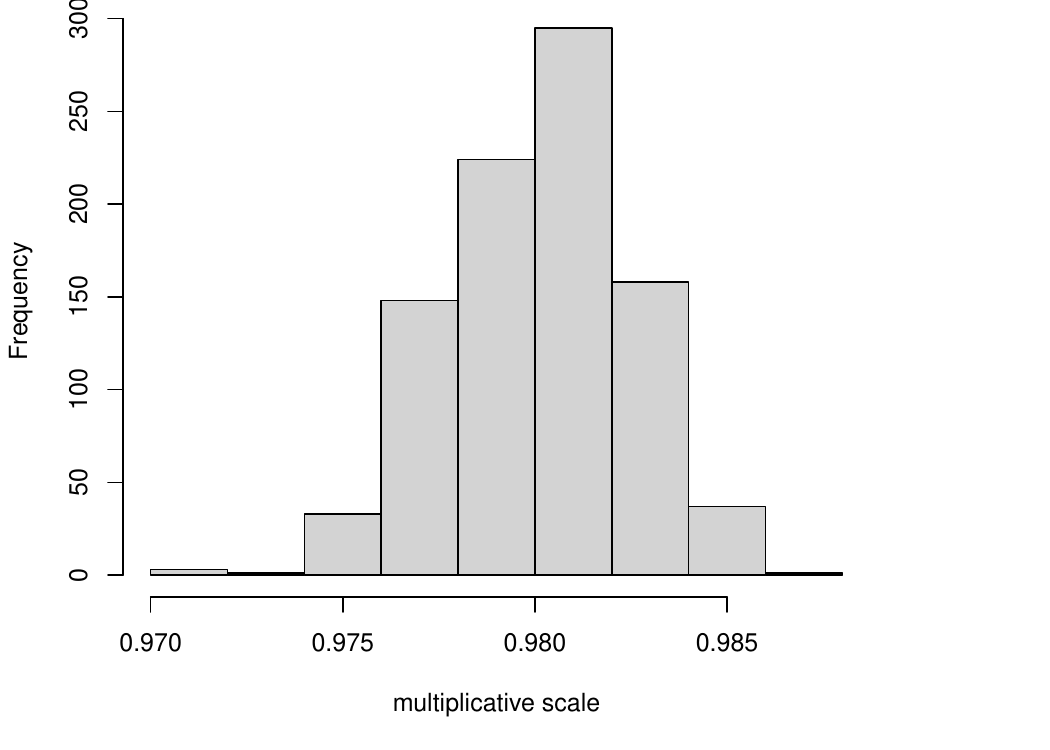}
\caption{Discrepancy (observed - estimated) between observed satellite data and
the simulator output at estimated model parameters in 2009 (top left), 2010
(top right), and 2011 (bottom left). Posterior samples of a multiplicative
scale discrepancy (bottom right).
\label{f:ibex_disc}}
\end{figure}

Figure \ref{f:ibex_disc} displays the discrepancy between what the Scaled
Vecchia surrogate of our computer model outputs at the posterior mean estimated
by our framework for the model parameter $u$ and data collected from the IBEX
satellite in the years 2009-2011. It's clear that the computer model does not
sufficiently account for the behavior observed in the satellite data. But this
is not terribly surprising. Simulators are inherently biased, simply from not
being able to completely reconstruct the real process.
\citet{kennedyohagan2001} address this limitation by introducing a discrepancy
term in their canonical computer model calibration framework, a practice
followed by many other practitioners
\citep{higdoncalib2004,higdoncalib2008,Liu2009ModularizationIB}.

The KOH approach uses a GP for modeling of complex disparities between
simulator and reality. Due to our field response being counts, further
methodology needs to be developed to apply the same GP discrepancy methods to
IBEX, a point we leave to future work. However, we test the addition of a
simple, multiplicative discrepancy term $\delta$ to our original model,
modifying Eq.~(\ref{eq:pois_inv_bayes_mod}) as follows:
\begin{align}
Y^F &\sim \mathrm{Pois}(\lambda(X^F), e(X^F)) & \lambda(X) &= m(u, X) \nonumber
\label{eq:mod_with_disc} \\ \lambda(X^F) &= \lambda(X)*\delta &
\mathrm{log}(\delta) &\sim \mathcal{N}(0, 1) \nonumber \\ u &\sim \pi(u) &
\mbox{e.g.,} \quad u &\sim \mathrm{Unif}[0,1]^p.
\end{align}

The bottom right panel of Figure \ref{f:ibex_disc} shows posterior samples of
$\delta$. Mass is concentrated just below 1, indicating nearly zero effect from
the multiplicative discrepancy. Examining the disparities closer, it appears
that spatial dependence exists in whether the simulator overestimates or
underestimates the mean surface. For instance, the area of the sky map
corresponding to the ribbon shows some underestimation. By contrast, areas
outside of the ribbon are overestimated. A constant multiplicative scale
discrepancy is clearly insufficient. We believe that $\delta$ is estimated
around 1.0 because of these competing interests between different regimes in
the sky map. More advancements are needed to properly model this disparity. A
fully Bayesian framework could be developed to provide full uncertainty quantification, perhaps with the use of a deep Gaussian process to handle the
inherent nonstationarity in the discrepancy.

We recognize that this is not a definitive calibration for IBEX data and some
oversimplifications were made to fit our narrative and present a more
generalizable method. Future work could account for details such as the
Compton-Getting effect \citep{compton1935apparent}, spatial blurring resulting
from the IBEX-Hi point spread function \citep{funsten2009IBEXHiENA}, and
varying parameters within a GDF computer simulation. We see the results in this
paper as a first step in iteration with the theoretical physicists on our team
toward the development of a higher fidelity model which more faithfully
reproduces dynamics observed in the field.

\subsection*{Data Availability}

The authors confirm that data and code supporting the conclusions of this
research, along with reproductions of included figures and examples, can be
found at \url{https://github.com/lanl/ibex-bayesian-inverse}.

\subsection*{Acknowledgments}

RBG and SDB are grateful for funding from NSF CMMI 2152679. SDB, DO and LJB
were funded by Laboratory Directed Research and Development (LDRD) Project
20220107DR. This work has been approved for public release under
LA-UR-25-30539.

\bibliography{ibex_bayes_inv}
\bibliographystyle{jasa}

\appendix

\section{Additional empirical validation}

Here we provide some supplementary results to further validate the methodology
introduced in the main body of our manuscript and lay additional groundwork
for future research endeavors.

\subsection{Probability Integral Transforms}
\label{app:pits}

Section \ref{sec:synth_data} showcased the ability of our Poisson Bayesian
inverse problem framework to recover the ``true'' model parameters $u$ from
synthetic satellite data. In particular, we made the claim that both sky maps
in Figure \ref{f:ibex_synth_ex} (bottom left, top right) could have reasonably
generated the ENA counts ``observed'' in the top left plot of that figure. This
assertion was based on a purely visual assessment of the data and the estimated
underlying ENA rate. A more empirical method to validate our claim could be via
the probability integral transform
\citep[PIT;][]{gneiting2007probabilistic,diebold1997evaluating,dawid1984present}.

PIT values, frequently visualized in PIT histograms, are used to evaluate
predictions and forecasts from a statistical model. In short, the PIT compares
observed data to the CDFs of predictive distributions. If observations came
from the predicted distributions, PIT values should be distributed uniformally
between 0 and 1. For discrete data such as our Poisson counts, PIT values will
not be uniformally distributed, due to the step function nature of the CDF. A
randomized PIT is used in these scenarios
\citep{gneiting2014probabilistic,czado2009predictive}, as we will do for the
IBEX data.

\begin{figure}[ht!]
\centering
\includegraphics[scale=0.5,trim=0 0 0 0,clip=TRUE]{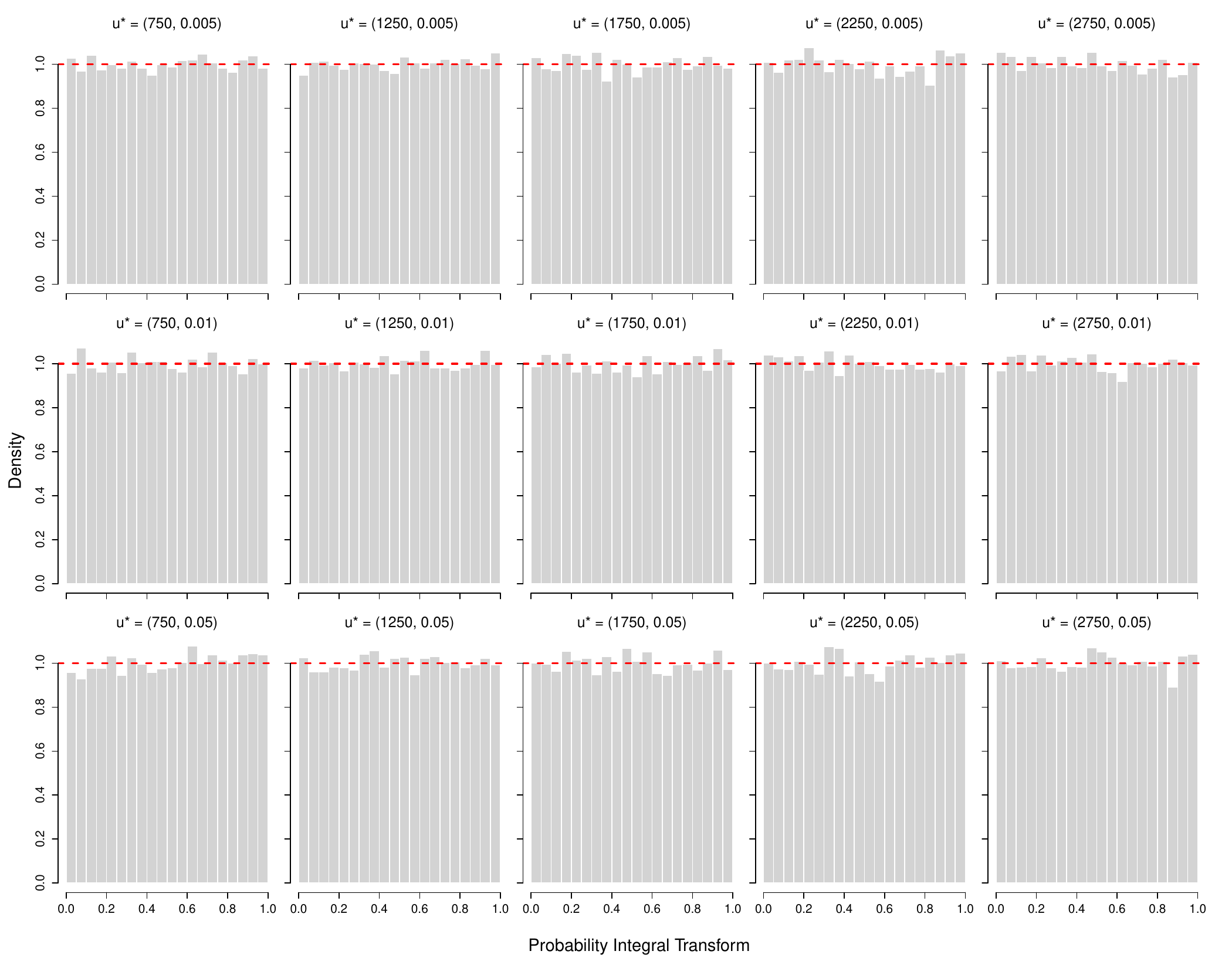}
\caption{Probability integral transform (PIT) histograms for 15 sets of
estimated sky maps and their corresponding synthetic ``observed'' datasets. Red
dashed lines are drawn at 1 to show where we'd expect the histograms to
approach as the number of draws goes to infinity.
\label{f:sim_pit_hist}}
\end{figure}

Figure \ref{f:sim_pit_hist} displays PIT histograms for comparing synthetic
counts to predicted ENA rates computed via a surrogate for our computer model
$m(\cdot,\cdot)$ at estimated model parameter $\hat{u}$. To keep it short, here
we show only 15 of the 36 cases from Figure \ref{f:synth_estimates_all}. Titles
of each plot indicate the true model parameter $u^\star$ that was used to
sample the synthetic data. If the data ``observed'' comes from a Poisson
distribution with a mean surface $\lambda(X) = \hat{m}(\hat{u}, X)$, then we'd
expect the PIT histogram to be distributed evenly between 0 and 1. This is
certainly the case for all 15 histograms shown.  No siginifcant departure from
a Uniform(0,1) is observed. Therefore, we can add further empirical backing to
our visual assessment that estimates from our framework reasonably explain the
data ``observed.''

\begin{figure}[ht!]
\centering
\includegraphics[scale=0.5,trim=0 0 0 0,clip=TRUE]{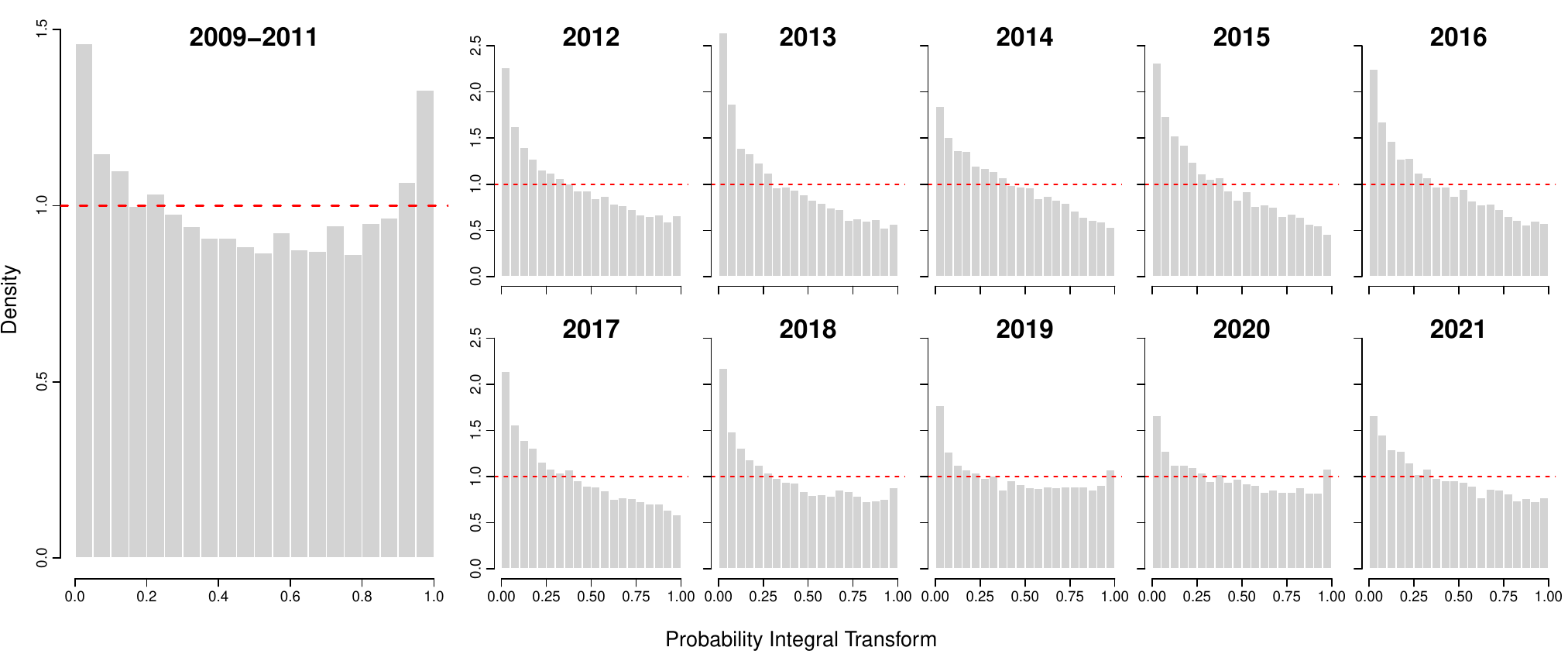}
\caption{Probability integral transform (PIT) histograms comparing ENA counts
collected by IBEX and predicted sky map from estimated model parameters. The
very left panel displays the PIT histogram for estimates from combined years
2009-2011. PIT histograms for individual years 2012-2021 are shown in the two
rows on the right. U-shaped PIT histograms indicate overdispersion.
Right-skewed PIT histograms indicate consistent overprediction.
\label{f:real_pit_hist}}
\end{figure}

In contrast to the results in Figure \ref{f:ibex_synth_ex} (and more broadly in
Figure \ref{f:synth_estimates_all}), our Poisson Bayesian inverse problem
framework struggles to estimate a value for $u$ that produces a map of ENA
rates that is likely to have generated the actual data observed from the IBEX
satellite. This is due in large part to the insufficiencies of the IBEX
simulator. To illustrate, consider Figure \ref{f:real_pit_hist}. Here we
display PIT histograms produced by comparing IBEX satellite data to surrogate
predicted sky maps using estimates of $u$ from our framework. On the left we
show results when satellite data from years 2009-2011 is used as the field data
($X^F, Y^F$). On the right we show results when treating years individually
from 2012-2021 as ($X^F, Y^F$). These panels correspond to Figures
\ref{f:ibex_real_ex} and \ref{f:real_estimates_all}, respectively.

The departure from uniformity is clear. For years 2009-2011, we get a
crater-shaped histogram, indicating that the predicted ENA rate surface does
not sufficiently account for the extreme observations in the satellite data.
That is, the observed data fell too frequently in the tails of our predicted
distribution. This is consistent with the spatially-varying discrepancy we
observed in Figure \ref{f:ibex_disc}. For years 2012-2021, we see a different,
though consistent, behavior. Although individual years vary, each PIT histogram
is right-skewed. This means that with the estimated $u$, the output from the
IBEX computer model, or the surrogate fit on the simulator output, is
systematically overpredicting the ENA rates for the years 2012-2021.

\subsection{Multiplicative Scale Discrepancy}
\label{app:scale_disc}

Equation \ref{eq:mod_with_disc} introduces a simple, multiplicative term
$\delta$ in an attempt to account for the gap between the predicted ENA rates
and ENA counts observed by IBEX. Unsurprisingly, we discovered that this
multiplicative scale was insufficient to account for the spatially-varying,
nonstationary discrepancy present in the satellite data. However, we gave no
evidence to suggest that, if indeed there only existed a multiplicative
discrepancy, our updated model would be able to correctly identify it.
Therefore, in order to verify that our scale discrepancy actually works, we
conducted an experiment on simulated data that fall within that model class.

First, we select one sky map from the simulator output corresponding to one of
the $n =66$ $u$-values and set this as $\lambda^\star$. Next, we both scale
$\lambda^\star$ by $\delta$ and add in the background rate, $\lambda_b$. Then,
we generate synthetic satellite data via counts drawn from a Poisson, $Y_i^F
\sim \mathrm{Pois}(\lambda_i^\star * \delta + \lambda_{b_i}, e_i)$ for given
exposure times $e_i$. Finally, we run Algorithm \ref{alg:gibbs}, modified
slightly to sample for a multiplicative scale discrepancy, and calculate a
posterior mean and 95\% credible intervals for $\delta$. We repeat this process
for 10 different values of $\delta$.

\begin{figure}[ht!]
\centering
\includegraphics[scale=0.65,trim=0 0 0 0,clip=TRUE]{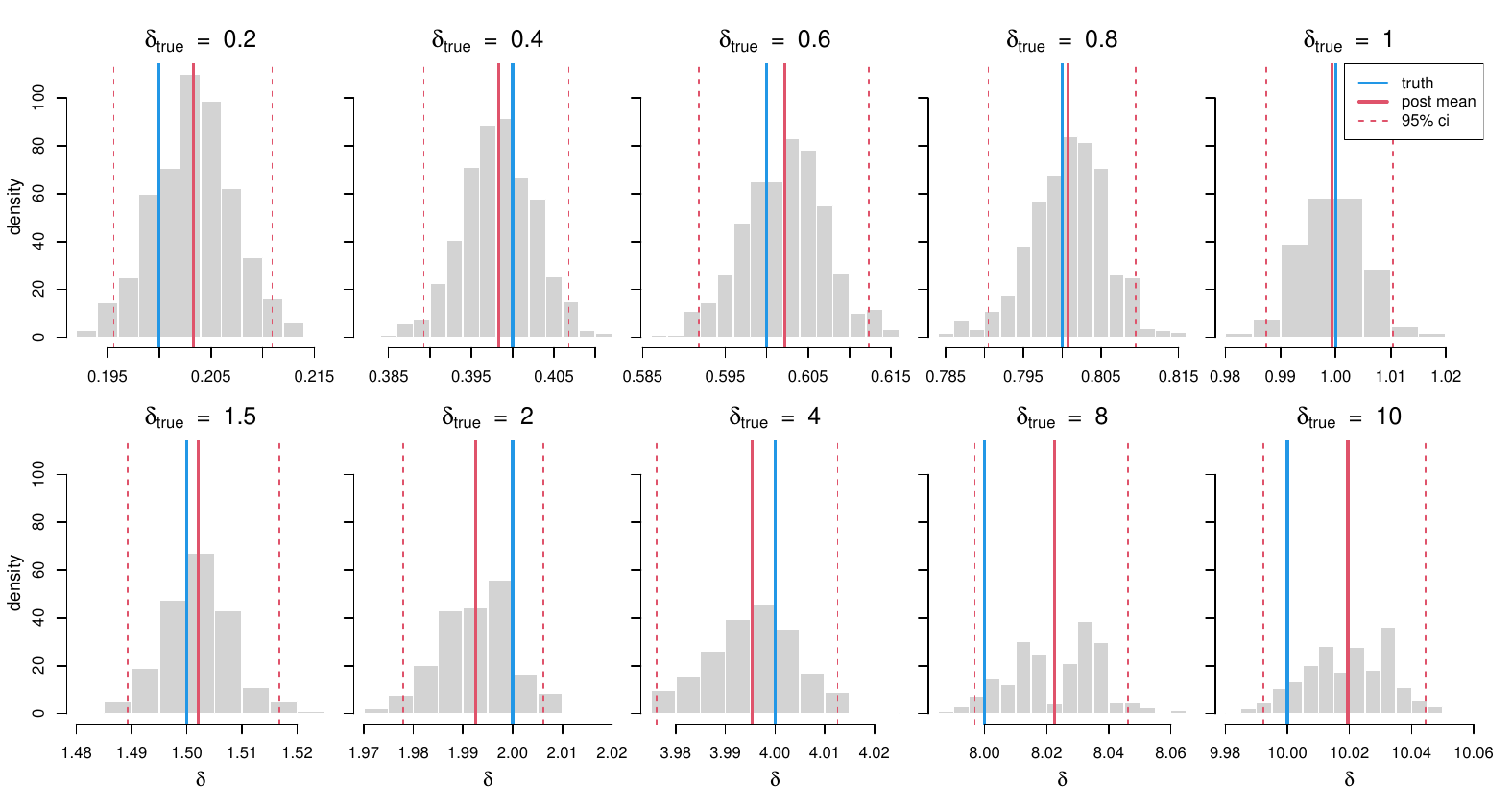}
\caption{Posterior samples of a multiplicative scale discrepancy, $\delta$. Each
panel corresponds to a Metropolis sampler run on synthetic data artificially
scaled by separate known constants ($\delta_{true}$). True values marked in
blue. Posterior means and 95\% credible intervals shown in red.
\label{f:scale_disc_samples}}
\end{figure}

Figure \ref{f:scale_disc_samples} shows histograms of posterior samples of
$\delta$ as estimated densities. We restrict the $y$-axis limits to be the
same, which causes different bar heights among the plots due to the variation
in spread of the posterior samples on the $x$-axes. In each of the 10 test
cases, our framework is able to recover the true value of $\delta$. Although
not shown here, it's important to mention that in addition to successfully
estimating $\delta$, our framework also recovered the true $u$-values used to
generate the synthetic data at the nominal 95\% rate. Therefore, we can be
confident that had the disparity between the IBEX simulator and the IBEX
satellite data been in the form of a multiplicative scale, our method would be
able to identity it. Alas, we look forward to future work exploring more
sophisticated discrepancy models and better capturing the mismatch between
simulator and reality.

\end{document}